\documentclass[12pt]{article}

\usepackage[utf8]{inputenc}
\usepackage[T1]{fontenc}
\usepackage[height=22.5cm,width=16.6cm]{geometry}

\usepackage{graphicx}
\usepackage{xcolor}
\usepackage{hyperref}

\usepackage{amsmath}
\usepackage{amsfonts}
\usepackage{amssymb}
\usepackage{amsthm}
\usepackage{bm}
\usepackage{comment}

\numberwithin{equation}{section}

\theoremstyle{definition}

\def\transpose{\mathsf{T}}

\def\Z{\mathbb{Z}}
\def\Q{\mathbb{Q}}
\def\R{\mathbb{R}}
\def\C{\mathbb{C}}

\def\cI{\mathcal{I}}
\def\cJ{\mathcal{J}}
\def\cN{\mathcal{N}}
\def\id{\mathrm{id}}

\def\End{\mathrm{End}}
\def\Hdg{\mathrm{Hdg}}

\def\Span{\mathrm{Span}}

\begin{document}
%

\begin{titlepage}
%
 
\vskip 3cm
\begin{center}
  
{\large \bf Towards Hodge Theoretic Characterizations \\ of 2d Rational SCFTs: II}
 
\vskip 1.2cm
  
Masaki Okada and Taizan Watari
  
\vskip 0.4cm
  
{\it 
    Kavli Institute for the Physics and Mathematics of the Universe (WPI), 
    University of Tokyo, Kashiwa-no-ha 5-1-5, 277-8583, Japan
}
 
\vskip 1.5cm
    
\abstract{A characterization of rational superconformal field theories (SCFTs) on 1+1 dimensions with Ricci-flat K\"{a}hler targets was proposed by S.\ Gukov and C.\ Vafa in terms of the Hodge structure of the target space.
The article [arXiv:2205.10299] refined this idea and extracted a set of necessary conditions for a $T^4$-target $\cN=(1,1)$ SCFT to be rational; only a partial effort was made, however, to study whether it also constitutes a sufficient condition.
It turns out that the set of conditions in [arXiv:2205.10299] is not sufficient, and that it becomes a set of necessary and sufficient conditions by adding one more condition in the case of $T^4$.
The Strominger--Yau--Zaslow fibration in the mirror correspondence plays an essential role there.
At the end, we also propose a refined version of Gukov--Vafa's idea for general Ricci-flat K\"{a}hler target spaces.
} 
\end{center}
\end{titlepage}

\tableofcontents

\begin{flushright}
\end{flushright}

\section{Introduction}
\label{sec:Intro}

A family of $(M, G)$ of a manifold $M$ of certain topological 
type that admits a Ricci-flat K\"{a}hler metric $G$ gives rise 
to a family of ${\cal N}=(1,1)$ superconformal field theories (SCFTs) on 
1+1 dimensions. Which points in the moduli space of those SCFTs 
correspond to rational CFTs? This article addresses that question. 

Based on observations by G. Moore \cite{Moore:1998zu, Moore:1998pn} 
and on examples in Gepner constructions, Gukov and Vafa proposed the following idea \cite{Gukov:2002nw}: 
\begin{quote}
The SCFTs that are rational may be characterized by CM-type complex 
structure of $M$ and CM-type complex structure of the mirror $W$ of $M$.
\end{quote}
Being {\it CM-type} is a property of complex structure of a manifold 
(or of Hodge structure on a cohomology group) that generalizes the 
notion of complex multiplication of $M=T^2$.
There are in fact a couple of subtle conceptual issues to be resolved 
when one wants to see if the characterization above works on practical 
cases; arbitrariness in the choice of complex structure (when $M$ is a 
complex torus, K3 surface, or a hyperk\"{a}hler manifold) is one of them. 
There is also a concrete example 
in Meng Chen's paper \cite{Chen:2005gm} that appears to be a counter example, 
when we read the characterization above naively. So, the characterizations (necessary and sufficient conditions) need to be stated in a way the conceptual issues are resolved, and Meng Chen's example does not satisfy them. 
Ref. \cite{Kidambi:2022wvh} did both, and presented a set of characterization
conditions that are at least necessary in the case $M = T^4$. 
One of the key observations in \cite{Kidambi:2022wvh} was that 
(i) the complex structure of $M$ should be such that $M$ is regarded 
as an algebraic variety than a complex analytic manifold, and 
(ii) the K\"{a}hler form of $M$ should be in the cone generated 
by divisors (the conditions 1 and 2(b) in \S 2), for the corresponding 
SCFT to be rational; without the observation (ii), one would still 
be haunted by Meng Chen's example of non-rational SCFTs. 

In this article, we continue along the line of \cite{Kidambi:2022wvh}. 
By adding one more condition to the set of conditions 
in \cite{Kidambi:2022wvh}, we obtain---for the family of 
$M=T^4$---a set of conditions that is necessary 
{\it and sufficient} for the SCFTs to be rational. 
The Strominger--Yau--Zaslow (SYZ) fibration in the mirror 
correspondence \cite{Strominger:1996it} turns out to play an essential role 
in the one condition added in this article (the condition 8 
in \S \ref{sec:one-more}). The triple of complex multiplication, 
mirror symmetry and algebraicity of the target space is the 
key to the characterization of rational SCFTs. The necessary 
and sufficient conditions for $M=T^2$ and $T^4$ are stated 
in a language that makes sense for a family with $M$ that 
is self-mirror (Conjecture 1 (self-mirror) in \S \ref{sec:towards});  
we do not have a proof for the cases beyond $M=T^2$ or $T^4$. 
As a wild speculation, a couple of trial versions of the 
characterization conditions are presented also for the cases with 
$M$ that is not self-mirror (Conj. 2 (general) in \S \ref{sec:towards}). 
Busy readers might just have a look at Theorem and Conjectures 
in section \ref{sec:towards}.

We will use the same notations and conventions as 
in \cite{Kidambi:2022wvh}, whenever possible. 
Review materials in \cite{Kidambi:2022wvh} as well as 
explanations on notations and conventions may be of some 
use while reading this article. We did not try very much 
to make this article self-contained, because this article is 
an outright continuation of \cite{Kidambi:2022wvh}. 
Various lecture notes and theses on the theory of complex 
multiplication and abelian varieties are available online 
(e.g., \cite{milnecm}); they will be useful when following the 
arguments in this article line by line. 

\section{Status Summary}
\label{sec:status-sum}

Let $(M; G, B)$ be a set of data that consists of 
a real $2n$-dimensional manifold $M$, a Riemannian 
metric $G$ on $M$ that is Ricci-flat and can be 
K\"{a}hler for some appropriate choice of a complex 
structure $I$, and a closed 2-form $B$ on $M$.  
The non-linear sigma model construction determines 
an $\cN=(1,1)$ SCFT for the data $(M; G, B)$. 
Reference \cite{Kidambi:2022wvh} used $M= T^4$ as 
a test case, and extracted several necessary conditions 
on the data $(M; G, B)$ for the SCFT to be rational. 

\vspace{5mm}

{\bf Thm.~5.8 of \cite{Kidambi:2022wvh}:}
The following is a list of necessary conditions 
in the case of $M= \R^4/\Z^{\oplus 4} = T^4$
for the $\cN=(1,1)$ SCFT to be rational 
(so, we should read $n=2$ here). 
\begin{enumerate}
  \item there exists a polarizable complex structure $I$ on $M$, with which $G$ is compatible and $(M,G,I)$ becomes K\"{a}hler, and $B^{\rm transc}=0$.
\end{enumerate}
For such a complex structure $I$ ($M_I$ is meant to be the complex manifold $(M, I)$), 
\begin{enumerate}
\setcounter{enumi}{1}
  \item the horizontal and vertical simple level-$n$ rational Hodge substructures satisfy the following 
  conditions:
  \begin{enumerate}
    \item The level-$n$ simple Hodge substructure on $[H^n(M;\Q)]_{\ell=n}$ by $I$ is of CM-type, where the CM field ${\rm End}([H^n(M;\Q)]_{\ell =n})^{\rm Hdg}$ is denoted by $K'$. 
    \item There exists a $[K':\Q]$-dimensional vector subspace of $A(M_I)\otimes \Q$ denoted by $T_M^v\otimes \Q$ on which a simple level-$n$ rational Hodge structure of weight-$n$ can be introduced, with the polarization
    \begin{align}
    H^\mathrm{even}(M;\Q)&\times H^\mathrm{even}(M;\Q)\ni(\psi,\chi)\nonumber\\
    &\longmapsto (-1)^{\frac{n(n-1)}{2}}\int_{M}\left(\sum_{k=0}^n(-1)^k\Pi_{2k}\psi\right)\wedge\chi\in\Q
    \label{eq:vert-poln}
    \end{align}
    where $\Pi_{2k}$ is the projection $H^\ast(M;\Q)\to H^{2k}(M;\Q)$.
    Its Hodge $(n,0)$ component is generated by $\mho := e^{2^{-1}(B+i\omega)}$, where $\omega = 2^{-1}G(I-,-)$ is the K\"{a}hler form, and this polarized rational Hodge structure is of CM-type, with the endomorphism field $K'$.
    \item There is an isomorphism of polarized rational Hodge structure of weight-$n$ between the vertical and horizontal simple level-$n$ components $T^v_M\otimes\Q$ and $[H^n(M;\Q)]_{\ell=n}$. 
  \end{enumerate}
  \item The rational Hodge substructures other than the weight-$n$ level-$n$ 
  components satisfy the following conditions: 
  \begin{enumerate}
    \item All other rational polarizable Hodge substructures on $H^k(M;\Q)$ 
    by $I$ are also of CM-type. 
    \item There is a filtration $W^\bullet_v$ on $H^*(M;\Q)$ so that the data $(\rho_{\rm spin}(h_{\omega,B}), W^\bullet_v)$ introduces a generalized rational Hodge structure on $H^*(M;\Q)$, so that
    \begin{enumerate}
      \item $T^v_M\otimes \Q \subset W^n_v$, and 
      \item the rational Hodge structures on $W^k_v/W^{k+2}_v$ is of CM-type for all $k$, and the one for $k=n$ is polarized by\footnote{
\label{fn:well-def-of-3(b)ii}
It is implied here that $W_v^n\subset H^{\mathrm{even}}(M;\Q)$ and that the subspace $W_v^{n+2} \subset W_v^n$ is orthogonal to any element 
in $W_v^n$ under the pairing (\ref{eq:vert-poln}).  
} the pairing (\ref{eq:vert-poln}).
    \end{enumerate}
  \end{enumerate}
\end{enumerate}
Furthermore, 
\begin{enumerate}
\setcounter{enumi}{3}
  \item there is a geometric SYZ-mirror\footnote{
A complex structure $I$ specifies one weight-1 operator $J$ in the 
left moving sector and one weight-1 operator $\tilde{J}$ in the
right moving sector. In the meanwhile, any T-duality / mirror 
correspondence is supposed to be an isomorphism of ${\cal N}=(1,1)$ 
SCFTs. It is non-trivial whether the isomorphism image of $J$ and 
$\tilde{J}$ can be interpreted by some mirror complex structure 
${}^\exists I_\circ$. Only the SYZ mirrors that pass this test are called 
geometric SYZ mirrors in this article. 
See \cite{VanEnckevort:2003qc} for more information. 
} %
  to the $\cN=(2,2)$ SCFT for the data $(M; G, B; I)$, and 
  \item (weak) at least for {\it some} of such geometric SYZ-mirrors, the 
  filtration $g^*(W_{h\circ})$ on $H^*(M;\Q)$ satisfies the properties of 
  $W_v^\bullet$ in the conditions 3, 6 and 7. 
%
\end{enumerate}
There is one more condition that makes sense only for a family of $(M;G,B)$ 
that is self-SYZ-mirror (as in the case of $M=T^{2n}$ and K3):
\begin{enumerate}
\setcounter{enumi}{5}
  \item there is a one-to-one correspondence between the simple rational horizontal Hodge substructures on $(W^k_h/W^{k+2}_h)$ and vertical Hodge substructures on $(W^k_v/W^{k+2}_v)$ so that there are Hodge isomorphisms.
\end{enumerate}
Finally, here is one more condition that is stated in a language applicable 
only to $M = T^{2n}$: 
\begin{enumerate}
\setcounter{enumi}{6}
  \item 
   the Hodge isomorphism in the condition 6, when chosen appropriately, can be 
    interpreted as a combination of an isogeny and a mirror map of D-brane charges. 
\end{enumerate}
One may wonder what happens if the condition 5(weak) above 
is replaced by 
\begin{enumerate}
\setcounter{enumi}{4}
\item (strong) For {\it any} geometric-SYZ mirror, the filtration 
$g^*(W_{h\circ})$ satisfies the property of $W_v^\bullet$ in the conditions 
3, 6 and 7. 
\end{enumerate}
Although the condition 5 of Thm.~5.8 of \cite{Kidambi:2022wvh} is read 
as 5(weak), one may read the derivation in Ref.\ \cite{Kidambi:2022wvh} 
and will realize that the condition 5 holds in the stronger version here
in the case of $M=T^4$ when the $\cN=(1,1)$ SCFT for $(M=T^4; G, B)$ is rational.
All the notations and concepts here are either explained or introduced 
already in \cite{Kidambi:2022wvh}. 

\vspace{5mm}

Let us first leave a few words about how to read the statements of 
Thm.~5.8 above. Reference \cite{Kidambi:2022wvh} worked on the cases 
where $M=T^4$ and extracted the properties of the data $(M; G,B)$ 
when the $\cN=(1,1)$ SCFT for $(M;G,B)$ is rational;
the properties were stated, however, in languages that make 
sense also for other Ricci-flat K\"{a}hler manifolds 
$M$. It was for this reason that the way the conditions are listed up 
is not optimized\footnote{
\label{fn:isog-Hdg-isom}
For example, an isogeny in the condition 7 implies a Hodge isomorphism in the condition 6 for $k=1$, from which Hodge isomorphisms in the condition 6 for all other $k$ follow, in the cases of $M=T^{2n}$.

Note that existence of an isogeny $T^{2n}_\circ\to T^{2n}$ is equivalent to existence of a Hodge isomorphism $H^1(T^{2n};\Q)\to H^1(T^{2n}_\circ;\Q)$ or of a Hodge isomorphism $H_1(T^{2n}_\circ;\Q)\to H_1(T^{2n};\Q)$; the former follows from the latter because
any Hodge isomorphism  $H_1(T^{2n}_\circ;\Q)\to H_1(T^{2n};\Q)$ can be multiplied 
by an appropriate integer to be a linear map $H_1(T^{2n}_\circ;\Z)\to H_1(T^{2n};\Z)$.
} %
for the case of $M=T^4$, or for the cases of $M=T^{2n}$.
The theorem statement above was not only a confirmed one for the $M=T^4$ 
case, but also meant to be a trial version for other Ricci-flat 
K\"{a}hler manifolds. 

\vspace{5mm}

The conditions 1--7 on a set of data $(M; G, B)_{M=T^4}$ are necessary 
conditions for the corresponding $\cN=(1,1)$ SCFT to be rational. 
It is a natural question whether the conditions 1--7 on a set of 
data $(M;G,B)$ are also {\it sufficient} conditions for the $\cN=(1,1)$ SCFT 
to be rational. This converse problem remains to be an open question 
even in the case $M=T^4$; Ref.\ \cite{Kidambi:2022wvh} made a partial 
attempt by imposing just the conditions 1--3 on $(M; G,B)_{M=T^4}$,
and found that some of such data correspond to $\cN=(1,1)$ SCFTs that 
are {\it not} rational. So, just the set of conditions 1--3 alone does
not constitute a sufficient condition for the $\cN=(1,1)$ SCFT to be rational. 

Therefore, there are two things we wish to do beyond
the work of \cite{Kidambi:2022wvh}.
One is (i) the converse problem: examine whether the set of all the conditions 1--7 constitutes a sufficient condition for the $\cN=(1,1)$ SCFT to be rational.
The other is (ii) to investigate relations among the conditions; some of the conditions might follow from others.
There are two motivations to think about (ii).
As a preparation for the converse problem (i), we aim to reduce the redundancy among the conditions 1--7 and simplify the problem.
As a preparation for application to more general Ricci-flat targets, 
we also aim to make the trial version of the statement as simple as possible.

We start working on both (i) and (ii) in the next section for the 
case of $M=T^4$. But there are a few quick comments on (ii) that we can make here. 

Whether the condition 2 (a) implies the condition 3 (a) or not: not enough mathematics is known at this moment.  
Although Rmk.\ 5.3 of \cite{Kidambi:2022wvh}  referred to this open question 
for abelian surfaces,
it now turns out that the answer is yes \cite{Okada:2023a}.
It is still an open question, however, for general Ricci-flat K\"{a}hler manifolds.
So, we choose to retain the condition 3 (a) at this moment 
in a set of sufficient conditions along with 
the condition 2 (a). Just as a reminder, the condition 2 (a) is placed 
before the condition 3 (a) in the statement because the weight-$n$ level-$n$
component $[H^n(M;\Q)]_{\ell =n}$ is always present and unique for a Ricci-flat 
K\"{a}hler manifold $M$ that is not necessarily a complex torus. 

As for the condition 7 (with 5(weak) or 5(strong)), the authors realized 
that the condition can be derived easily from the conditions 3, 4, 
5(weak/strong) and 6, when $M=T^{2n}$.
This is because those other conditions already imply that there is a Hodge isomorphism 
between $H^1(T^{2n}_I;\Q) \cong W_h^1/W_h^3$ and 
$W_v^1/W_v^3 \cong_{{\rm by~}g^*} W_{h\circ}^1/W_{h\circ}^3 \cong H^1(T^{2n}_{I\circ};\Q)$; see footnote \ref{fn:isog-Hdg-isom} for more. 
This observation (footnote \ref{fn:isog-Hdg-isom} in particular) is elementary;
the authors should have been aware while working for Ref. \cite{Kidambi:2022wvh}. 

Conversely, only the condition 7+5(weak/strong) is enough\footnote{
The detailed argument is as follows.
As in footnote \ref{fn:isog-Hdg-isom}, the condition 6+5 immediately follows, and the condition 3(b)ii+5 also follows from 3(a).
As for the condition 3(b)i+5, since $\mho\in H^\ast(T^{2n};\C)$ is the generator of the unique charge-$n$ eigenspace of the U(1) action $\rho_\mathrm{spin}(h_{\omega,B})$, it must be in $W_v^n\otimes\C=g^\ast(W_{h\circ}^n\otimes\C)$.
Furthermore, since the Hodge structure on $T_M^v\otimes\Q$ is of CM-type with the 
endomorphism field $K'$ as in the condition 2(b), we know that (a) 
$\mho \in W_v^n\otimes \C$ is in fact contained within $W_v^n \otimes \tau'_{(n,0)}(K')$ 
for a certain embedding $\tau'_{(n,0)}:K'\hookrightarrow\overline{\Q}$, and also 
that (b) $T_M^v\otimes\C$ is generated by its Galois conjugates 
$\{\mho^\sigma\mid\sigma\in\mathrm{Gal}(\overline{\Q}/\Q)\}$ 
(cf. \cite[Lemma\ A.11]{Kidambi:2022wvh}).
Since all the Galois conjugates $\mho^\sigma$ are in $W_v^n\otimes\C$, 
 $T_M^v\otimes\Q\subset W_v^n$, that is, the condition 3(b)i+5 is satisfied.
} %
for the conditions 3(b), 5(weak/strong) and 6, when a set of data $(M;G,B)_{M=T^{2n}}$ satisfies the conditions 1--3(a) and 4.
It still makes sense to write down the conditions by using the Hodge isomorphisms 
of cohomology groups as in the conditions 3(b)--6 than in terms of geometry 
(in the condition 7), because isogenies make sense only for the case of $M=T^{2n}$. 

\section{Relations among the Conditions}
\label{sec:relations}

In this section, we show that the conditions 4 and 7+5(strong) on a set of data $(M; G, B)$ (and hence 3(b), 6+5(strong) also) follow from the 
conditions 1--3(a), when $M=T^4$. So, as long as the cases with $M=T^4$ 
are concerned, we may retain only the conditions 1--3(a) and drop all 
others from Thm.\ 5.8 of \cite{Kidambi:2022wvh}, and yet no information is lost.
As is evident from the following discussions, however, we will resort to a case-by-case 
analysis for $M=T^4$ in this section, so we must say that it is not clear 
whether the conditions 3(b)--7 follow from the conditions 1--3(a)
even when $M= T^{2n}$ with $n>2$. 

To prove that the conditions 4 and 7+5(strong) are satisfied when the conditions 1--3(a)
are imposed on a set of data $(T^4; G, B)$, we can take advantage 
of the analysis of \cite[\S5.3]{Kidambi:2022wvh}. There, a set of data $(T^4; G, B; I)$ 
satisfying the conditions 1--3(a) is classified into one of the four cases, (A), (A'), (B and C), 
based on the complex structure $(T^4, I)$, to get started; when $(T^4, I)$ is 
in the case (A), then it was shown in \S5.3.3 of \cite{Kidambi:2022wvh}
that the set of data $(T^4; G, B)$ gives rise to a rational $\cN=(1,1)$ SCFT, 
so the data also satisfy the conditions 3(b)--7. When $(T^4, I)$ is 
either in the case (A') or in the case (B, C), however, the complexified 
K\"{a}hler parameter $(B+i\omega)$ on $(T^4, I)$ is not necessarily that 
for a rational $\cN=(1,1)$ SCFT.

Here, we give a minimum summary of the results from \cite{Kidambi:2022wvh} 
and a reminder of some notations on the cases (A') and (B, C) for those who 
do not want to go through \cite{Kidambi:2022wvh}. For a positive rational 
number $d$, $\sqrt{d}$ stands for a real positive number. For a negative 
rational number $p$, $\sqrt{p} = i \sqrt{-p}$ is a pure imaginary complex 
number in the upper half complex plane throughout this article.
Lemmas A.11 and A.12 of \cite{Kidambi:2022wvh} on CM-type rational 
Hodge structures are elementary, to the extent that math lecture notes 
do not explain at length, but are still used in this article frequently, 
sometimes even without a mention. 

An abelian surface $T^4_I$ in the case (A') is isogenous to the product 
$E_1\times E_2$ of a pair ($E_1$ and $E_2$) of mutually non-isogenous CM 
elliptic curves, whose complex structure parameters are 
$\tau_1\in\Q(\sqrt{p_1})$ and $\tau_2\in\Q(\sqrt{p_2})$, 
respectively 
(i.e., one may choose $p_1$, $p_2$ to be negative square-free integers 
such that $p_1\not\in p_2(\Q^\times)^2$).
In this case, we have 
\begin{align}
\End(H^1(E_i;\Q))^\Hdg\cong\Q(\sqrt{p_i}),\quad\End(H^1(T^4;\Q))^\Hdg\cong\Q(\sqrt{p_1})\oplus\Q(\sqrt{p_2}).
\end{align}
We can take the rational basis $\{\hat{\alpha}^i,\hat{\beta}_i\}$ of $H^1(E_i;\Q)$ and the complex coordinate $z^i$ of $E_i$ so that $dz^i=\hat{\alpha}^i+\sqrt{p_i}\hat{\beta}_i$.
The action of $\xi_i\in\Q(\sqrt{p_i})$ as an element of $\End(H^1(E_i;\Q))^\Hdg$ is just multiplying $dz^i$ by $\xi_i$.
The rational basis of $H^1(T^4;\Q)$ obtained by pulling back $\hat{\alpha}^1,\hat{\beta}_1,\hat{\alpha}^2,\hat{\beta}_2$ of $H^1(E_1\times E_2;\Q)$ by an isogeny $T^4\to E_1\times E_2$ is also denoted by $\hat{\alpha}^1,\hat{\beta}_1,\hat{\alpha}^2,\hat{\beta}_2$. 

When a set of data $(T^4;G,B;I)$ in the case (A') satisfies the conditions 1--3(a), the complexified K\"{a}hler parameter $(B+i\omega)/2$ 
is of the form of either one of 
\begin{align}
(B+i\omega)/2 = \left(A + C \sqrt{p_1} \right) \hat{\alpha}^1\hat{\beta}_1
    + \left(\widetilde{A} + \widetilde{C} \sqrt{p_2} \right)
       \hat{\alpha}^2\hat{\beta}_2,       \qquad 
    A, \widetilde{A} \in \Q, \;  C, \widetilde{C} \in \Q_{\neq 0},
   \label{eq:Aprime-B+iomega-RCFT}
\end{align}
or 
\begin{align}
(B+i\omega)/2 = \left(A + C \sqrt{p_2} \right) \hat{\alpha}^1\hat{\beta}_1
    + \left(\widetilde{A} + \widetilde{C} \sqrt{p_1} \right)
       \hat{\alpha}^2\hat{\beta}_2,       \qquad 
    A, \widetilde{A} \in \Q, \;  C, \widetilde{C} \in \Q_{\neq 0} .
    \label{eq:Aprime-B+iomega-nonRCFT}
\end{align}
A set of data $(T^4; G, B, I)$ that falls into the 
case (\ref{eq:Aprime-B+iomega-RCFT}) corresponds to a rational 
$\cN=(1,1)$ SCFT, and a set of data that falls into 
the case (\ref{eq:Aprime-B+iomega-nonRCFT}) to an $\cN=(1,1)$ 
SCFT that is not rational. 
Derivation is found in \S5.3.2 of \cite{Kidambi:2022wvh}. 

An abelian surface $T^4_I$ in the cases (B, C) is a simple abelian variety 
of CM-type. The endomorphism algebra $\End(H^1(T^4;\Q))^\Hdg$ is a 
degree-4 CM field, which has a structure 
\begin{align}
K\cong\Q[x,y]/(y^2-d,x^2-p-qy) 
\end{align}
where $d>1$ is a positive square-free integer, and $p,q\in\Q$ are subject to 
the conditions $p<0$, $q\neq0$ and $d':=p^2-q^2d>0$.
The case (B) is when $K$ is Galois over $\Q$, and the case (C) 
when $K$ is not; the distinction between them is not important in this article, 
however. The four embeddings $K\hookrightarrow\overline{\Q}$ are denoted 
by $\tau_{\pm\pm}$, where
\begin{align}
\tau_{\pm,\ast}(y)=\pm\sqrt{d},\qquad & \tau_{\pm+}(x)=\sqrt{\pm}:=i\sqrt{-p\mp q\sqrt{d}}\ ,\\
& \tau_{\pm-}(x)=-\sqrt{\pm}=-i\sqrt{-p\mp q\sqrt{d}}\ .
\end{align}

The reflex field $K^r$ of the CM-pair $(K,\{\tau_{++},\tau_{-+}\})$ 
in the cases (B, C) has the structure 
\begin{align}
K^r\cong\Q[y',\xi^r]/((y')^2-d',(\xi^r)^2-2p+2y').
\end{align}
The four embeddings $K^r\hookrightarrow\overline{\Q}$ are denoted by $\tau^r_{\pm\pm}$, where
\begin{align}
\tau^r_{\pm,\ast}(y')=\pm\sqrt{d'}=\mp\sqrt{+}\sqrt{-},\qquad & \tau^r_{\pm+}(\xi^r)=\sqrt{+}\pm\sqrt{-}\ ,\\
& \tau^r_{\pm-}(\xi^r)=-(\sqrt{+}\pm\sqrt{-}).
\end{align}
The reflex field $K^r$ is also a degree-4 CM field.

When a set of data $(T^4;G,B;I)$ in the cases (B, C) satisfies the conditions 1--3(a), the complexified K\"{a}hler parameter $(B+i\omega)/2$
is of the form 
\begin{align}
 (B+i\omega)/2 = Z_1 e_1 + Z_2 e_2,     \label{eq:BC-B+iomega}     
\end{align}
where
\begin{align}
& e_1:=\hat{\alpha}^1\hat{\beta}_1+d\hat{\alpha}^2\hat{\beta}_2,\quad e_2:=\hat{\alpha}^1\hat{\beta}_2-\hat{\beta}_1\hat{\alpha}^2,\\
& Z_1 := \tau^r_{++}\left( A + \frac{C'}{2} \xi^r
      + \frac{D'}{2d} \frac{2ad}{\xi^r} \right), \quad 
Z_2 := \tau^r_{++} \left( \widetilde{A} \pm \frac{D'}{2} \xi^r 
        \pm \frac{C'}{2}\frac{2qd}{\xi^r} \right), \\
& (A, \widetilde{A}, C', D' \in \Q, (C',D')\neq(0,0))
\end{align}
as was shown in \S5.3.1 of \cite{Kidambi:2022wvh}. A set of data 
that falls into the case (\ref{eq:BC-B+iomega};$+$) corresponds to 
a rational $\cN=(1,1)$ SCFT. The $\cN=(1,1)$ SCFT that corresponds to a 
set of data in the case (\ref{eq:BC-B+iomega};$-$), on the other hand, 
is not rational.  

We already know that the conditions 3(b)--7 are satisfied in the cases where
the complex structure $(T^4, I)$ and the complexified K\"{a}hler 
parameter $(B+i\omega)/2$ of the data $(T^4; G, B; I)$ are in the 
case (A')--(\ref{eq:Aprime-B+iomega-RCFT}) and  
case (B, C)--(\ref{eq:BC-B+iomega};$+$), because the conditions 3(b)--7
are satisfied by the set of data $(T^4; G, B)$ for a rational $\cN=(1,1)$ 
SCFT \cite{Kidambi:2022wvh}. In the rest of this section, we will show 
that the conditions 3(b)--7 are satisfied also in the cases 
(A')--(\ref{eq:Aprime-B+iomega-nonRCFT}) and (B, C)--(\ref{eq:BC-B+iomega};$-$).

\subsection{The Case (A')}
\label{subsec:Aprime-1-7}
 
As a first step, let us list up all the possible  
geometric SYZ-mirrors of the $\cN=(2,2)$ SCFT for a 
set of data $(T^4; G, B; I)$ in the case 
of (A')--(\ref{eq:Aprime-B+iomega-nonRCFT}).  
By finding a non-empty list, we prove that the condition 4
follows automatically in the case (A')--(\ref{eq:Aprime-B+iomega-nonRCFT}); 
the full list of geometric SYZ-mirrors is used to prove the 
condition 7+5(strong) later in this subsection \ref{subsec:Aprime-1-7}.

Recall that, for the T-duality taken along 1-cycles specified by a rank-$n$ primitive subgroup $\Gamma_f\subset H_1(T^{2n};\Z)$, the corresponding geometric SYZ-mirror exists if and only if the following conditions are satisfied \cite[Prop.~8]{VanEnckevort:2003qc}:
\begin{align}
\omega|_{\Gamma_f\otimes\R}=0,\qquad B|_{\Gamma_f\otimes\R}=0.
\label{eq:SYZ-condition}
\end{align}
To obtain the list of such $\Gamma_f$'s, we first list up all the rank-$n$ subspaces $\Gamma_{f\Q}\subset H_1(T^{2n};\Q)$ satisfying $\omega|_{\Gamma_{f\Q}\otimes\R}=0$ and $B|_{\Gamma_{f\Q}\otimes\R}=0$.
Then $\Gamma_{f\Q}\cap H_1(T^{2n};\Z)$ for these $\Gamma_{f\Q}$ constitute the list of all $\Gamma_f$ satisfying (\ref{eq:SYZ-condition}).

Let $\Gamma_f=\Span_\Q\{c',d'\}$ be such a rank-$(n=2)$ subspace of $H_1(T^{2n=4};\Q)$.
We parameterize the generators by
\begin{align}
c':=\ &c_1\alpha_1+c_2\beta^1+c_3\alpha_2+c_4\beta^2,\\
d':=\ &d_1\alpha_1+d_2\beta^1+d_3\alpha_2+d_4\beta^2,\\
&c_1,\ldots,c_4,d_1,\ldots,d_4\in\Q\,.
\end{align}
The above condition $\omega|_{\Gamma_{f\Q}\otimes\R}=B|_{\Gamma_{f\Q}\otimes\R}=0$ for (A')--(\ref{eq:Aprime-B+iomega-nonRCFT}) is then equivalent to
\begin{align}
&c_1:c_2=d_1:d_2 \quad \text{and} \quad c_3:c_4=d_3:d_4,
\end{align}
where we used the positive volume condition $C, \widetilde{C} \neq 0$. 
Therefore, we can reorganize the basis of $\Gamma_{f\Q}$ so that
\begin{align}
\Gamma_{f\Q}=\Span_\Q\{c,d\},\quad &c:=c_1\alpha_1+c_2\beta^1,\quad d:=c_3\alpha_2+c_4\beta^2,
\label{eq:Aprime-GammafQ}\\
&c_1,\ldots,c_4\in\Q, (c_1,c_2),(c_3,c_4)\neq(0,0).
\label{eq:Aprime-GammafQ-prm}
\end{align}
The $\Gamma_{f\Q}$'s of our interest are exhausted by those in the form of (\ref{eq:Aprime-GammafQ}) with the parameters scanned as in (\ref{eq:Aprime-GammafQ-prm}).\footnote{
Precisely the same analysis can be repeated for the case (A')--(\ref{eq:Aprime-B+iomega-RCFT}); one then finds that the $\Gamma_{f\Q}$'s are parameterized precisely as in (\ref{eq:Aprime-GammafQ}) and (\ref{eq:Aprime-GammafQ-prm}).
In Ref.~\cite[\S4.2]{Kidambi:2022wvh}, only the choice $c_2=c_4=0$ was presented as an example of geometric SYZ-mirrors.}

Once such an $(n=2)$-dimensional subspace $\Gamma_{f\Q} \subset H_1(T^{2n=4};\Q)$ is chosen, the \mbox{rank-$n$} primitive subgroup $\Gamma_f := \Gamma_{f\Q}\cap H_1(T^{2n};\Z)$ has been determined.
For any such $\Gamma_f$, there is a complement rank-$n$ subgroup $\Gamma_b \subset H_1(T^{2n};\Z)$ so that $\Gamma_f \oplus \Gamma_b \cong H_1(T^{2n};\Z) \cong \Z^{\oplus 2n}$; there are infinitely many different ways to choose $\Gamma_b$ for a given $\Gamma_f$.
Every choice of such $(\Gamma_f,\Gamma_b)$ specifies a geometric SYZ-mirror; $\Gamma_f$ and $\Gamma_b$ are meant to be the 1-cycles along which T-duality is taken and not taken, respectively.
This is the end of the process of listing up all the $(\Gamma_f,\Gamma_b)$'s for geometric SYZ mirrors.

Although the condition 4 has been verified, let us also write down some more explicit information on cohomologies of $T^4$ and $T^4_\circ$ for later convenience.
For a given $n$-dimensional subspace $\Gamma_{f\Q} \subset H_1(T^{2n};\Q)$,
an $n$-dimensional subspace $\Gamma_{b\Q}^\vee \subset H^1(T^{2n};\Q)$ 
is specified as those that vanish on $\Gamma_{f\Q}$. In other words, 
they are cohomologies on $T^{2n}$ that are pulled back from the 
cohomologies on the base by the SYZ $T^n$ fibration. This subspace 
$\Gamma_{b\Q}^\vee$ is generated by 
\begin{align}
  \hat{e} := -c_2 \hat{\alpha}^1 + c_1 \hat{\beta}_1, \qquad 
  \hat{f} := -c_4 \hat{\alpha}^2 + c_3 \hat{\beta}_2. 
\end{align}
Note that we may use $\{c,d,\hat{e},\hat{f}\}$ as a basis of
\begin{align}
H^1(T^4_\circ;\Q) \cong \Gamma_{f\Q} \oplus \Gamma_{b\Q}^\vee = {\rm Span}_\Q \left\{ c, d, \hat{e}, \hat{f} \right\};
\label{eq:Aprime-H^1-mirror}
\end{align}
T-dualized 1-cycles of $T^4$ are also regarded\footnote{
\label{fn:construction-of-mirror}
The isomorphism $\Gamma_{f\Q}\oplus\Gamma_{b\Q}^\vee\cong H^1(T^4_\circ;\Q)$ is a part of the map $g:H_1(T^{2n};\Z)\oplus H^1(T^{2n};\Z)\to H_1(T^{2n}_\circ;\Z)\oplus H^1(T^{2n}_\circ;\Z)$ of the winding and Kaluza--Klein charges which identify states on both sides of the T-dual with the same masses.
The notation ``$g$'' of the map is omitted in (\ref{eq:Aprime-H^1-mirror}) and also in other parts of 
this article except the appendix \ref{app:isog}. 
For the spinor representation of this $g$, which is 
$H^*(T^{2n};\Q) \longrightarrow H^*(T^{2n}_\circ;\Q)$, however, 
the notation "$g$" is recycled and used in the main text. 
} %
as 1-cocycles of the mirror $T^4_\circ$, so we abused notations a little bit
here. 

Let us choose a basis $\{e, f\}$ of $\Gamma_b \otimes \Q$ so that $\{e, f\}$ is dual 
to the basis $\{\hat{e}, \hat{f} \}$ of $\Gamma_{b\Q}^\vee \subset H^1(T^4;\Q)$. 
Then\footnote{
\label{fn:choice-of-Gammab}
If we are to choose $e\in (c_1^2+c_2^2)^{-1}e'+\Gamma_{f\Q}$ and $f\in (c_3^2+c_4^2)^{-1}f'+\Gamma_{f\Q}$ arbitrarily, however, there is no guarantee that the primitive subgroup $\Gamma_b':=\Span_\Q\{e,f\}\cap H_1(T^4;\Z)$ satisfies $\Gamma_f\oplus\Gamma_b'=H_1(T^4;\Z)$.
} %
$e \in (c_1^2+c_2^2)^{-1} e' + \Gamma_{f\Q}$ and $f \in (c_3^2+c_4^2)^{-1} f' + \Gamma_{f\Q}$, where 
\begin{align}
   e' :=  -c_2\alpha_1 + c_1\beta^1, \qquad   f' := -c_4 \alpha_2+c_3\beta^2 .    
\end{align}
The choice of a basis $\{c,d,e,f\}$ of $H_1(T^4;\Q)$ also determines its dual basis of $H^1(T^4;\Q)$, 
which are denoted by $\{\hat{c}, \hat{d}, \hat{e}, \hat{f} \}$. 
For a given $\Gamma_{f\Q}$ (and hence a given $\Gamma_f$), 
we may choose $\{ \hat{c}' + \Gamma_{b\Q}^\vee, \hat{d}' + \Gamma_{b\Q}^\vee \}$ as 
a basis of the coset vector space $H^1(T^4;\Q)/\Gamma_{b\Q}^\vee$, with 
\begin{align}
  \hat{c}' :=  c_1\hat{\alpha}^1 + c_2\hat{\beta}_1 , \qquad 
  \hat{d}' :=  c_3\hat{\alpha}^2 + c_4\hat{\beta}_2  
  \label{eq:Aprime-e'f'chat'dhat'}
\end{align}
independent of a choice of $\Gamma_b$, and yet $\hat{c} \in (c_1^2+c_2^2)^{-1} \hat{c}' + \Gamma_{b\Q}^\vee$
and $\hat{d} \in (c_3^2+c_4^2)^{-1} \hat{d}' + \Gamma_{b\Q}^\vee$. It often happens in the analysis in this article 
that the basis $\{ \hat{e}, \hat{f}, \hat{c}', \hat{d}' \}$ of $H^1(T^4;\Q)$ is more convenient than $\{ \hat{e}, \hat{f}, \hat{c}, \hat{d} \}$
(as we only have to refer to a choice of $\Gamma_f$, not of $\Gamma_b$). 

Now that the condition 4 has been verified, let us move on to show that the condition 7+5(strong) is satisfied.
That is to verify\footnote{
Although the following proof for the existence of an isogeny exploits properties very specific to the case (A'), which allows us to make the proof simple, we can also find isogenies by direct calculations of the rational Hodge structures just like we will do for the cases (B, C) in section \ref{subsec:BC-1-7}.
} %
the existence of an isogeny $T^4_I\to T^4_{I\circ}$ for all the geometric SYZ-mirrors.

Recall that $T^4_I$ in the case (A') is isogenous to the product $E_1\times E_2$ of CM elliptic curves $E_1$ and $E_2$, whose complex structure parameters are $\tau_1\in\Q(\sqrt{p_1})$ and $\tau_2\in\Q(\sqrt{p_2})$, respectively.
In addition, in the case of (A')--(\ref{eq:Aprime-B+iomega-nonRCFT}), their complexified K\"{a}hler parameters are $\rho_1\in \Q(\sqrt{p_2})$ and $\rho_2\in \Q(\sqrt{p_1})$, respectively.
For any $\Gamma_{f\Q}$ in (\ref{eq:Aprime-GammafQ}), as shown 
in the appendix \ref{app:isog}, the geometric SYZ-mirror 
$T^4_{I\circ}$ is isogenous to $E_1^\circ\times E_2^\circ$, 
where $E_1^\circ$ is the T-dual of $E_1$ along $c\in H_1(E_1;\Q)$, 
and $E_2^\circ$ that of $E_2$ along $d\in H_1(E_2;\Q)$.
Here, the complex structure parameters $\tau_1^\circ$ of $E_1^\circ$ and $\tau_2^\circ$ of $E_2^\circ$ are in the imaginary quadratic fields $\Q(\rho_1)=\Q(\sqrt{p_2})=\Q(\tau_2)$ and $\Q(\rho_2)=\Q(\sqrt{p_1})=\Q(\tau_1)$, respectively.
Therefore, there exist isogenies $E_1 \rightarrow E_2^\circ$ and $E_2\rightarrow E_1^\circ$, and hence also an isogeny $T^4_I \rightarrow T^4_{I\circ}$.

This argument for the existence of an isogeny holds for any choice of geometric SYZ-mirror, so we have verified the condition 7+5(strong).
This completes the proof of the conditions 4 and 7+5(strong) (and hence 3(b)--7) for the case (A')--(\ref{eq:Aprime-B+iomega-nonRCFT}).

\subsection{The Case (B, C)}
\label{subsec:BC-1-7}

Let us move on to the proof of the conditions 4 and 7+5(strong) for the case (B,C)--(\ref{eq:BC-B+iomega};$-$).
The outline of the proof is the same as in section \ref{subsec:Aprime-1-7}. 

Our first task in this section \ref{subsec:BC-1-7} is to list up all the rank-$n$ subspaces $\Gamma_{f\Q}\subset H_1(T^{2n};\Q)$ satisfying $\omega|_{\Gamma_{f\Q}\otimes\R}=0$ and $B|_{\Gamma_{f\Q}\otimes\R}=0$.
Let $\Gamma_f=\Span_\Q\{c',d'\}$ be such a rank-$n$ subspace of $H_1(T^{2n=4};\Q)$.
The condition $\omega|_{\Gamma_{f\Q}\otimes\R}=B|_{\Gamma_{f\Q}\otimes\R}=0$ is then equivalent to
\begin{align}
e_1(c',d')=e_2(c',d')=0;
\label{eq:BC-SYZcond-e1e2}
\end{align}
we have used the fact that the volume of $T^4$ must be positive ($\omega\neq0$, i.e.\ $(C',D')\neq(0,0)$), and 
that the integer $d$ is square-free.

Let us rewrite the condition (\ref{eq:BC-SYZcond-e1e2}) in a way useful for later analysis, by 
parametrizing the generators $c',d'$ of $\Gamma_{f\Q}$ as
\begin{align}
c':=\ &c_1\alpha_1+c_2\beta^1+c_3\alpha_2+c_4\beta^2,\\
d':=\ &d_1\alpha_1+d_2\beta^1+d_3\alpha_2+d_4\beta^2,\\
&c_1,\ldots,c_4,d_1,\ldots,d_4\in\Q\,.
\end{align}
To proceed, note that there are linear combinations of $e_1$ and $e_2$ that are decomposable:
\begin{align}
e_1+\sqrt{d}e_2=(\hat{\alpha}^1+\sqrt{d}\hat{\alpha}^2)(\hat{\beta}_1+\sqrt{d}\hat{\beta}_2)=\hat{\gamma}^1\hat{\delta}_1,\\
e_1-\sqrt{d}e_2=(\hat{\alpha}^1-\sqrt{d}\hat{\alpha}^2)(\hat{\beta}_1-\sqrt{d}\hat{\beta}_2)=\hat{\gamma}^2\hat{\delta}_2,
\end{align}
where 
\begin{align}
\hat{\gamma}^1:=\hat{\alpha}^1+\sqrt{d}\hat{\alpha}^2,\quad & \hat{\delta}_1:=\hat{\beta}_1+\sqrt{d}\hat{\beta}_2,\\
\hat{\gamma}^2:=\hat{\alpha}^1-\sqrt{d}\hat{\alpha}^2,\quad & \hat{\delta}_2:=\hat{\beta}_1-\sqrt{d}\hat{\beta}_2.
\end{align}
The condition (\ref{eq:BC-SYZcond-e1e2}) is therefore equivalent to 
$\hat{\gamma}^1\hat{\delta}_1(c',d')=\hat{\gamma}^2\hat{\delta}_2(c',d')=0$.
This translates to 
\begin{align}
&\left\{\begin{array}{l}
(c_1+\sqrt{d}c_3):(c_2+\sqrt{d}c_4)=(d_1+\sqrt{d}d_3):(d_2+\sqrt{d}d_4)\\
(c_1-\sqrt{d}c_3):(c_2-\sqrt{d}c_4)=(d_1-\sqrt{d}d_3):(d_2-\sqrt{d}d_4)
\end{array}\right..
\label{eq:BC-SYZcond-cd}
\end{align}

The generator $d'$ for a given $c'$ is therefore constrained by the relation (\ref{eq:BC-SYZcond-cd}); let us see how, by translating the relation (\ref{eq:BC-SYZcond-cd}) further. 
An immediate consequence of the relation (\ref{eq:BC-SYZcond-cd}) is that there exist $k_1,k_2\in\C$ 
such that $d'$ can be rewritten as
\begin{align}
d'=\,&\frac{1}{2}\Bigl((c_1+\sqrt{d}c_3)k_1+(c_1-\sqrt{d}c_3)k_2\Bigr)\alpha_1+\frac{1}{2\sqrt{d}}\Bigl((c_1+\sqrt{d}c_3)k_1-(c_1-\sqrt{d}c_3)k_2\Bigr)\alpha_2
\nonumber\\
&+\frac{1}{2}\Bigl((c_2+\sqrt{d}c_4)k_1+(c_2-\sqrt{d}c_4)k_2\Bigr)\beta^1+\frac{1}{2\sqrt{d}}\Bigl((c_2+\sqrt{d}c_4)k_1-(c_2-\sqrt{d}c_4)k_2\Bigr)\beta^2.
\label{eq:BC-d'-rational}
\end{align}
Since the generator $d'$ is an element of $H_1(T^4;\Q)$, the coefficients in (\ref{eq:BC-d'-rational}) must be 
rational numbers.
It follows from this fact and some easy calculation that $k_1,k_2\in\Q(\sqrt{d})$, and moreover
\begin{align}
&k_1=a+b\sqrt{d},\ k_2=a-b\sqrt{d}\\
&(a,b\in\Q,(a,b)\neq(0,0)).
\end{align}
So, the generator $d'$ has to be of the form 
\begin{align}
d'=(ac_1+bdc_3)\alpha_1+(bc_1+ac_3)\alpha_2+(ac_2+bdc_4)\beta^1+(bc_2+ac_4)\beta^2
\label{eq:BC-d'-result}
\end{align}
for the conditions (\ref{eq:BC-SYZcond-e1e2}, \ref{eq:BC-SYZcond-cd}) to be satisfied. 
This (\ref{eq:BC-d'-result}) is also sufficient. 

This brings us to the conclusion\footnote{
Precisely the same analysis can be repeated for the case of (\ref{eq:BC-B+iomega};$+$); one then finds that the $\Gamma_{f\Q}$'s are parameterized precisely as in (\ref{eq:BC-GammafQ}).
In Ref.~\cite[\S4.2]{Kidambi:2022wvh}, only the choice $c_2=c_3=c_4=0$ was presented as an example 
of geometric SYZ-mirrors. 
} %
that the $\Gamma_{f\Q}$'s of our interest (the $n$ 1-cycles 
along which T-dualities are taken to be geometric SYZ-mirrors) 
are exhausted by those of the form\footnote{
apologies for our poor choice of notations: $d \in \Gamma_{f\Q}$
here and a positive integer $d$ that generates $\Q(\sqrt{d})$
} %
\begin{align}
\Gamma_{f\Q}=\Span_\Q\{c,d\},\quad &c:=c_1\alpha_1+c_2\beta^1+c_3\alpha_2+c_4\beta^2,
\nonumber\\
&d:=dc_3\alpha_1+dc_4\beta^1+c_1\alpha_2+c_2\beta^2,
\label{eq:BC-GammafQ}\\
&c_1,\ldots,c_4\in\Q, c_1^2+c_2^2+c_3^2+c_4^2\neq0
\label{eq:BC-GammafQ-prm}
\end{align}
 with the parameters scanned as in (\ref{eq:BC-GammafQ-prm}). The directions 
 $\Gamma_{b} \subset H_1(T^4;\Z)$ in which T-duality is not taken 
 are subject only to the condition that $\Gamma_f \oplus \Gamma_b \cong H_1(T^4;\Z)$. 

Now that we have found a non-empty list of geometric SYZ-mirrors, the condition 4 is verified.
For later convenience, however, let us leave some more explicit information on cohomologies 
of $T^4$ and $T^4_\circ$.

Think of a $\Gamma_{f\Q}$ with $(c_1,c_3)\neq(0,0)$, first, for simplicity. 
The $n$-dimensional subspace $\Gamma_{b\Q}^\vee\subset H^1(T^{2n};\Q)$ that vanishes on $\Gamma_{f\Q}$ is generated by 
\begin{align}
\hat{e}&:=(c_1^2-dc_3^2)\hat{\beta}_1-(c_1c_2-dc_3c_4)\hat{\alpha}^1-d(c_1c_4-c_2c_3)\hat{\alpha}^2,
\label{eq:BC-c_1c_3nonzero-ehat}\\
\hat{f}&:=(c_1^2-dc_3^2)\hat{\beta}_2-(c_1c_4-c_2c_3)\hat{\alpha}^1-(c_1c_2-dc_3c_4)\hat{\alpha}^2,
\label{eq:BC-c_1c_3nonzero-fhat}
\end{align}
and we may use $\{c,d,\hat{e},\hat{f}\}$ as a basis of
\begin{align}
H^1(T^4_\circ;\Q) \cong \Gamma_{f\Q} \oplus \Gamma_{b\Q}^\vee = \Span_\Q \left\{ c, d, \hat{e}, \hat{f} \right\};
\label{eq:BC-H^1-mirror}
\end{align}
T-dualized 1-cycles of $T^4$ are also regarded as 1-cocycles of the mirror $T^4_\circ$, so we abused notations a little bit here again. 

Let us choose a basis $\{e, f\}$ of $\Gamma_b \otimes \Q$ so that $\{e, f\}$ is dual to the basis 
$\{ \hat{e}, \hat{f} \}$ of $\Gamma_{b\Q}^\vee \subset H^1(T^4;\Q)$. 
Then\footnote{
The remark in footnote \ref{fn:choice-of-Gammab} also holds true here.
} %
$e \in (c_1^2-dc_3^2)^{-1} e' + \Gamma_{f\Q}$ and $f \in (c_1^2-dc_3^2)^{-1} f' + \Gamma_{f\Q}$, where 
\begin{align}
    e' := \beta^1, \qquad  f' := \beta^2. 
\end{align}
The choice of a basis $\{ c,d,e, f\}$ of $H_1(T^4;\Q)$ determines its dual basis of $H^1(T^4;\Q)$, 
denoted by $\{ \hat{c}, \hat{d}, \hat{e}, \hat{f}\}$. While $\{ \hat{c}, \hat{d} \}$ generates 
the subspace $\Gamma_{f\Q}^\vee \subset H^1(T^4;\Q)$ that vanishes on $\Gamma_{b\Q}$, 
the coset space $H^1(T^4;\Q)/\Gamma_{b\Q}^\vee$ has a basis $\{\hat{c}' + \Gamma_{b\Q}^\vee, 
\hat{d}' +\Gamma_{b\Q}^\vee\}$, where 
\begin{align}
\hat{c}' :=  c_1\hat{\alpha}^1 - dc_3\hat{\alpha}^2 , \qquad 
\hat{d}' :=  - c_3\hat{\alpha}^1 + c_1\hat{\alpha}^2 ,
\label{eq:BC-c_1c_3nonzero-e'f'chat'dhat'}
\end{align}
and $\hat{c} \in (c_1^2-dc_3^2)^{-1} \hat{c}' + \Gamma_{b\Q}^\vee$ and 
$\hat{d} \in (c_1^2-dc_3^2)^{-1} \hat{d}' + \Gamma_{b\Q}^\vee$. 

Think of a general $\Gamma_{f\Q}$ now, when there is no guarantee that $(c_1,c_3) \neq 0$. 
Because $c\neq 0$, however, $(c_2,c_4) \neq 0$ then. So the $\Gamma_{f\Q}$'s that are not covered 
by the argument above is covered by $\Gamma_{f\Q}$'s where $(c_2,c_4) \neq 0$. 
Suppose, now, that $(c_2,c_4)\neq 0$. The same discussion as above holds for this case, when 
\begin{align}
\hat{e}& \; :=(c_2^2-dc_4^2)\hat{\alpha}^1-(c_1c_2-dc_3c_4)\hat{\beta}_1-d(c_2c_3-c_1c_4)\hat{\beta}_2,
\label{eq:BC-c_2c_4nonzero-ehat}\\
\hat{f}& \; :=(c_2^2-dc_4^2)\hat{\alpha}^2-(c_2c_3-c_1c_4)\hat{\beta}_1-(c_1c_2-dc_3c_4)\hat{\beta}_2,
\label{eq:BC-c_2c_4nonzero-fhat} 
\end{align}
and 
\begin{align}
e' & \; := \alpha_1, \quad f' := \alpha_2, \quad
\hat{c}' := c_2\hat{\beta}_1 - dc_4\hat{\beta}_2, \quad 
\hat{d}' :=  - c_4\hat{\beta}_1 + c_2\hat{\beta}_2;
\end{align}
$\hat{c} \in (c_2^2-dc_4^2)^{-1} \hat{c}' + \Gamma_{b\Q}^\vee$ and 
$\hat{d} \in (c_2^2-dc_4^2)^{-1} \hat{d}' + \Gamma_{b\Q}^\vee$ now. 
We could write down an expression for $\hat{e}$, $\hat{f}$ for a $\Gamma_{f\Q}$ with a fully general $c \neq 0$, it is much easier in the analysis in this article to work on the cases with $(c_1,c_3) \neq 0$, 
and on those with $(c_2,c_4) \neq 0$ because the expressions of those generators remain as simple as above then. 
In the rest of this article, we will work only on the cases with $(c_1,c_3) \neq 0$, and will not repeat 
the same discussions for the cases with $(c_2,c_4) \neq 0$. 

\vspace{5mm}

The remaining task is to verify the condition 7+5(strong): the existence of an isogeny $T^4\to T^4_\circ$ for all the geometric SYZ-mirrors.
So, we will show that $W_v^3$ ($\cong_\text{by $g^\ast$}H^3(T^4_\circ;\Q)$) is Hodge isomorphic to $H_1(T^4;\Q)$ in the following (cf. footnote \ref{fn:isog-Hdg-isom}).

The vector subspace $W_v^3$ in $H^*(T^4;\Q)$ is determined by pulling back
\begin{align}
W_{h\circ}^3=H^3(T^4_\circ;\Q)\cong\Span_\Q\{cd\hat{e},cd\hat{f},c\hat{e}\hat{f},d\hat{e}\hat{f}\}
\end{align}
by the map\footnote{
See \cite[\S 3, \S 9.3]{Golyshev:1998vzz} or \cite[footnote 13]{Kidambi:2022wvh}.
} of D-brane charges $g^\ast:H^\ast(T^4_\circ;\Q)\to H^\ast(T^4;\Q)$, which is 
\begin{align}
W_v^3=g^\ast(W_{h\circ}^3)=\ &\Span_\Q\{\hat{e},\hat{f},\hat{d}\hat{e}\hat{f},\hat{c}\hat{e}\hat{f}\}
\label{eq:BC-W3-Qbasis1}\\
=\ &\Span_\Q\{\hat{e},\hat{f},e_1\hat{f},\ e_2\hat{f}\}.
\label{eq:BC-W3-Qbasis2}
\end{align}
We might (or might not) use in the rest of this article that $e_1\hat{f} = -e_2 \hat{e}$ and $e_2\hat{f} = - e_1 \hat{e}/d$.

The vertical rational Hodge structure on $W^3_v$---equivalent 
to the rational Hodge structure on $H^3(T^4_\circ;\Q)$ 
with respect to the complex structure of the mirror---is 
determined by recalling the following fact:
its Hodge (1, 0) component is $\mho\wedge (\Gamma_{b\Q}^\vee \otimes \C)$, which has two generators 
\begin{align}
&\mho\hat{e}=\hat{e}+Z_1e_1\hat{e}+Z_2e_2\hat{e}\quad\text{and} \quad 
\mho\hat{f}=\hat{f}+Z_1e_1\hat{f}+Z_2e_2\hat{f}.
\label{eq:BC-Wv3-charge1-basis}
\end{align}
We claim that this rational Hodge structure on $W^3_v$ is 
of CM-type. An easy way to see this is to reorganize the 
two generators into the form 
where \cite[Lemma A.12]{Kidambi:2022wvh} can be used
 (ignore the sign choice\footnote{
The sign written below ($+$ in $\mp$ and $-$ in $\pm$) is for the 
case (B, C)--(\ref{eq:BC-B+iomega};$-$), while the sign written above 
is for the case (B, C)--(\ref{eq:BC-B+iomega};+).
} %
$-$ [resp. $+$] in $\mp$ [resp. $\pm$] here)\footnote{
The eigenbasis of the action of the CM field on $W^3_v=g^*(W^3_{h\circ})$ was worked out for 
both the case (B, C)--(\ref{eq:BC-B+iomega};+) 
and (B, C)--(\ref{eq:BC-B+iomega};$-$), 
for the choice $(c_1,c_2,c_3,c_4)=(1,0,0,0)$ already in \cite[eq.~(5.40)]{Kidambi:2022wvh}.
The present version is for a general $\Gamma_{f\Q}$ of a geometric SYZ mirror.  
} %
\begin{align}
\begin{pmatrix}
\mho\hat{e} & \mho\hat{f}
\end{pmatrix}
\begin{pmatrix}
1 & 1\\
\mp\sqrt{d} & \pm\sqrt{d}
\end{pmatrix}
=
\begin{pmatrix}
\hat{e} & \hat{f} & e_1\hat{f} & e_2\hat{f}
\end{pmatrix}
\begin{pmatrix}
1 & 1\\
\tau_{++}(\mp y) & \tau_{-+}(\mp y) \\
\tau_{++}(-\Xi_\pm) & \tau_{-+}(-\Xi_\pm)\\
\tau_{++}(\mp\Xi_\pm y) & \tau_{-+}(\mp\Xi_\pm y)
\end{pmatrix},
\label{eq:BC-W3-CM}
\end{align}
where
\begin{align}
\Xi_\pm:=\tilde{A}\pm Ay\pm D'x\pm C'xy\in K;
\label{eq:Xi}
\end{align}
it is now easy to see from the way the two elements 
of the basis in (\ref{eq:BC-W3-CM}) are related to 
a rational basis 
 (see, e.g., \cite[Lemma~A.12]{Kidambi:2022wvh}) that 
the weight-1 vertical rational Hodge structure on $W^3_v$ (horizontal rational Hodge structure on $H^3(T^4_\circ;\Q)$)
is of CM-type. The CM-pair is $(K, \{ \tau_{++}, \tau_{-+}\})$, 
which is identical to that of $H_1(T^4;\Q)$. 
This means that there is a Hodge isometry between $H_1(T^4;\Q)$ and $H_1(T^4_\circ;\Q)$. 

This is what we needed to show the existence of an isogeny $T^4\to T^4_\circ$. The argument holds for any choice of geometric SYZ-mirror, 
so the condition 7+5(strong) is now verified.
This completes the proof of the conditions 4 and 7+5(strong) (and hence 3(b)--7) for the case (B, C)--(\ref{eq:BC-B+iomega};$-$).

\section{One More Condition}
\label{sec:one-more}

The conditions (with 5(strong)) listed up in 
Thm.\ 5.8 of \cite{Kidambi:2022wvh} are necessary conditions 
for the $\cN=(1,1)$ SCFT of a set of data $(T^4; G, B)$ to 
be rational (as shown in \cite{Kidambi:2022wvh}), but fails 
to be a sufficient condition as a whole. This is because 
Ref.\ \cite{Kidambi:2022wvh} already exploited the 
conditions 1--3(a) and found that the $\cN=(1,1)$
SCFT is not rational in the cases of (A')--(\ref{eq:Aprime-B+iomega-nonRCFT})
and (B, C)--(\ref{eq:BC-B+iomega};$-$); we have now seen 
in the previous section that 
the properties 3(b), 4, 5(strong), 6 (and 7) follow from the conditions 1--3(a),
and hence are satisfied by those cases, although the SCFTs are not rational.

We claim here, and will prove in this section \ref{sec:one-more}, 
that one can add one more condition to the conditions 1--6 and turn 
them into a set of necessary and sufficient conditions on a set of data 
$(T^4; G, B)$ for the corresponding $\cN=(1,1)$ SCFT to be rational. 
There are two versions in doing so. One is to add to 
the conditions 1--6 with 5(weak); 
\begin{enumerate}
\setcounter{enumi}{7}
\item (weak) there exists a Hodge isomorphism 
$\phi^*: H^1(T^4;\Q) \rightarrow H^1(T^4_\circ;\Q)$ for {\it some} geometric 
SYZ-mirror satisfying the conditions 5(weak), 3 and 6 such that 
\begin{align}
\phi^*|_{\Gamma_{b\Q}^\vee} = {\rm id}_{\Gamma_{b\Q}^\vee}.
  \label{eq:cond-8}
\end{align}  
\end{enumerate}
The other is to add to the conditions 1--6 with 5(strong);
\begin{enumerate}
\setcounter{enumi}{7}
\item (strong) there exists a Hodge isomorphism 
$\phi^*: H^1(T^4;\Q) \rightarrow H^1(T^4_\circ;\Q)$ satisfying (\ref{eq:cond-8}) for {\it any} geometric 
SYZ-mirror.
\end{enumerate}
Obviously a set of data satisfying the condition 8(strong) satisfies 
the condition 8(weak); a set of data not satisfying the condition 8(weak) 
does not satisfy the condition 8(strong) either. 


In the rest of this section \ref{sec:one-more}, we will see 
that the condition 8(weak) is not satisfied by a set of data 
in the cases (A')--(\ref{eq:Aprime-B+iomega-nonRCFT}) 
and (B, C)--(\ref{eq:BC-B+iomega};$-$).
It will also turn out that the condition 8(strong) is satisfied 
in the cases (A')--(\ref{eq:Aprime-B+iomega-RCFT}), 
(B, C)--(\ref{eq:BC-B+iomega};$+$) and (A).
This means that the set of conditions 1, 2, 3(a) and 8 (either weak or strong is fine)
on a set of data $(T^4; G, B)$ is necessary and sufficient for the corresponding 
$\cN=(1,1)$ SCFT to be rational.

\subsection{The Case (A')}
\label{subsec:cond8-caseAprm}

First, let us show that the condition 8(weak) is not satisfied 
by a set of data $(T^4;G,B;I)$ in the case (A')--(\ref{eq:Aprime-B+iomega-nonRCFT}); we do so in a proof by contradiction.

Assume that a set of data $(T^4;G,B;I)$ in the case (A')--(\ref{eq:Aprime-B+iomega-nonRCFT}) satisfies the condition 8(weak); 
let $(\Gamma_f,\Gamma_b)$ be the choice for the geometric SYZ-mirror and $\phi^\ast:H^1(T^4;\Q)\to H^1(T^4_\circ;\Q)$ be the 
Hodge isomorphism in the condition.
We will write down explicitly (i) what is implied by the condition $\phi^*|_{\Gamma_{b\Q}^\vee} ={\rm id}_{\Gamma_{b\Q}^\vee}$ and 
(ii) what is implied by $\phi^\ast$ being a Hodge isomorphism, and then show that there cannot be such a $\phi^*$ satisfying 
both (i) and (ii). 

(i) The condition $\phi^\ast|_{\Gamma_{b\Q}^\vee}=\mathrm{id}_{\Gamma_{b\Q}^\vee}$ can be expressed in the form of 
\begin{align}
\phi^\ast
\begin{pmatrix}
\hat{c}' & \hat{d}' & \hat{e} & \hat{f}
\end{pmatrix}
=
\begin{pmatrix}
c & d & \hat{e} & \hat{f}
\end{pmatrix}
\left(\hspace{-5pt}\begin{array}{c|c}
r_{kl} & \hspace{-3pt}\begin{array}{c}0\\\bm{1}\end{array}
\end{array}\hspace{-6pt}\right),
\label{eq:Aprime-cond8-asEq}
\end{align}
by using the basis $\{ \hat{c}', \hat{d}', \hat{e}, \hat{f} \}$ of $H^1(T^4;\Q)$ and the basis $\{c,d,\hat{e},\hat{f}\}$ of $H^1(T^4_\circ;\Q)$ 
explained in (\ref{eq:Aprime-H^1-mirror}) and (\ref{eq:Aprime-e'f'chat'dhat'}), respectively. Here, 
 $(r_{kl})$ is a $4\times2$ $\Q$\hspace{1pt}-valued matrix.

Now, the condition (\ref{eq:Aprime-cond8-asEq}) implies that the holomorphic basis\footnote{
\label{fn:comment-normalization-dz}
The normalization factor $(c_1^2+c_2^2)$ and $(c_3^2+c_4^2)$ are inserted on the left hand side, 
only to make sure that $dz^1$ and $dz^2$ here have the same normalization as those in \cite{Kidambi:2022wvh}. 
They are not essential at all. 
} %
\begin{align}
\begin{pmatrix}
(c_1^2+c_2^2)dz^1 & (c_3^2+c_4^2)dz^2
\end{pmatrix}
& =
\begin{pmatrix}
\hat{c}' & \hat{d}' & \hat{e} & \hat{f}
\end{pmatrix}
\begin{pmatrix}
c_1+c_2\sqrt{p_1} & \\
& c_3+c_4\sqrt{p_2} \\
-c_2+c_1\sqrt{p_1} & \\
& -c_4+c_3\sqrt{p_2}
\end{pmatrix}
\end{align}
should be mapped by $\phi^\ast$ into
\begin{align}
\phi^\ast_\C
\begin{pmatrix}
(c_1^2+c_2^2)dz^1 & (c_3^2+c_4^2)dz^2
\end{pmatrix}
=
\begin{pmatrix}
c & d & \hat{e} & \hat{f}
\end{pmatrix}
\left(\hspace{-5pt}\begin{array}{c|c}
r_{kl} & \hspace{-3pt}\begin{array}{c}0\\\bm{1}\end{array}
\end{array}\hspace{-6pt}\right)
\begin{pmatrix}
c_1+c_2\sqrt{p_1} & \\
& c_3+c_4\sqrt{p_2} \\
-c_2+c_1\sqrt{p_1} & \\
& -c_4+c_3\sqrt{p_2}
\end{pmatrix}.
\label{eq:Aprime-isogeny-on-Q-basis}
\end{align}

(ii) Next, we claim that $\phi^*$'s being a Hodge isomorphism is equivalent to the presence of 
$\theta_1 \in \Q(\sqrt{p_1})^\times$ and $\theta_2 \in \Q(\sqrt{p_2})^\times$ such that 
\begin{align}
\phi^\ast_\C
\begin{pmatrix}
(c_1^2+c_2^2)dz^1 & (c_3^2+c_4^2)dz^2
\end{pmatrix}
& =
\begin{pmatrix}
dz^2_\circ & dz^1_\circ
\end{pmatrix}
\begin{pmatrix}
\theta_1 & \\
 & \theta_2
\end{pmatrix};
\label{eq:Aprime-isogeny-on-C-basis-preparation}
\end{align}
here, $dz^1_\circ$ and $dz^2_\circ$ are holomorphic 1-forms on $E_1^\circ$ and $E_2^\circ$ (pulled back to $T^4_\circ$), respectively, 
normalized so that they return a rational value for some rational element in $H_1(E_1^\circ;\Q)$ and $H_1(E_2^\circ;\Q)$, respectively.
This claim is based on the two following observations; first, there is an isogeny $\psi:T^4\sim E_1\times E_2 \leftarrow E_2^\circ\times E_1^\circ\sim T^4_\circ$ 
in the case (A')--(\ref{eq:Aprime-B+iomega-nonRCFT}), as we have seen in section \ref{subsec:Aprime-1-7}; this means that 
the map $\psi^*$ has the form of (\ref{eq:Aprime-isogeny-on-C-basis-preparation}) with $\theta_1, \theta_2$ replaced by 
some elements $\theta^\psi_1 \in \Q(\sqrt{p_1})^\times$ and $\theta^\psi_2 \in \Q(\sqrt{p_2})^\times$ for $\psi$.
Second, any Hodge isomorphism $\phi^*: H^1(T^4;\Q) \rightarrow H^1(T^4_\circ;\Q)$ can differ from $\psi^*$
only by\footnote{
Note that $\phi^* = (\phi^* \circ (\psi^*)^{-1}) \circ \psi^*$, and that 
$(\phi^* \circ (\psi^*)^{-1}) \in[\End(H^1(T^4_\circ;\Q))^\Hdg]^\times$.
} %
an additional action by $[{\rm End}(H^1(T^4_\circ;\Q))^{\rm Hdg}]^\times \cong  \Q(\sqrt{p_2})^\times\oplus\Q(\sqrt{p_1})^\times$.

Here, the holomorphic 1-forms on the mirror, $dz^1_\circ$ and $dz^2_\circ$, must be in the form of
\begin{align}
\begin{pmatrix}
\lambda'_1dz^2_\circ & \lambda'_2 dz^1_\circ
\end{pmatrix}
&=
\begin{pmatrix}
d & \hat{f} & c & \hat{e}
\end{pmatrix}
\begin{pmatrix}
1 & \\
\rho'_1 & \\
 & 1\\
 & \rho'_2
\end{pmatrix}
\end{align}
for\footnote{
\label{fn:mirror-dependence-of-rho}
Just as a reminder, these $\lambda'_1,\rho'_1,\lambda'_2,\rho'_2$ depend on the isogeny $T^4_{I\circ}\to E_1^\circ\times E_2^\circ$, whose pullback is used to identify $H^1(T^4_\circ;\Q)$ and $H^1(E_1^\circ;\Q)\oplus H^1(E_2^\circ;\Q)$, and hence depend on $(\Gamma_f,\Gamma_b)$ for the geometric SYZ-mirror we chose at the beginning of the proof.
} some $\lambda'_1, \rho'_1 \in \Q(\sqrt{p_1})^\times$ and 
$\lambda'_2, \rho'_2 \in \Q(\sqrt{p_1})^\times$.
Therefore, (\ref{eq:Aprime-isogeny-on-C-basis-preparation}) becomes
\begin{align}
\phi^\ast_\C
\begin{pmatrix}
(c_1^2+c_2^2)dz^1 & (c_3^2+c_4^2)dz^2
\end{pmatrix}
& =
\begin{pmatrix}
c & d & \hat{e} & \hat{f}
\end{pmatrix}
\begin{pmatrix}
 & 1\\
1 & \\
 & \rho'_2\\
\rho'_1  & 
\end{pmatrix}
\begin{pmatrix}
\theta'_1 & \\
 & \theta'_2
\end{pmatrix}, 
\label{eq:Aprime-isogeny-on-C-basis}
\end{align}
where $\theta'_1 := \theta_1/\lambda'_1 \in \Q(\sqrt{p_1})^\times$ and $\theta'_2:=\theta_2/\lambda'_2 \in \Q(\sqrt{p_2})^\times$.

\vspace{5mm}

Now that the properties (i) and (ii) in the condition 8 have been paraphrased as (\ref{eq:Aprime-isogeny-on-Q-basis}) and (\ref{eq:Aprime-isogeny-on-C-basis}) respectively, let us see that there is no common solution $(r_{kl},\theta'_i)$ to (\ref{eq:Aprime-isogeny-on-Q-basis}) and (\ref{eq:Aprime-isogeny-on-C-basis}) indeed, or equivalently, no solution to 
\begin{equation}
\left(\hspace{-5pt}\begin{array}{c|c}
r_{kl} & \hspace{-3pt}\begin{array}{c}0\\\bm{1}\end{array}
\end{array}\hspace{-6pt}\right)
\begin{pmatrix}
c_1+c_2\sqrt{p_1} & \\
& c_3+c_4\sqrt{p_2} \\
-c_2+c_1\sqrt{p_1} & \\
& -c_4+c_3\sqrt{p_2}
\end{pmatrix}
=
\begin{pmatrix}
 & 1\\
1 & \\
 & \rho'_2\\
\rho'_1 & 
\end{pmatrix}
\begin{pmatrix}
\theta'_1 & \\
 & \theta'_2
\end{pmatrix},
\label{eq:Aprime-isogeny-compare}
\end{equation}
for any $(c_\text{1--4},\rho'_{1,2})$. To see this, note that the both sides of (\ref{eq:Aprime-isogeny-compare}) 
are $4\times2$ matrices after multiplication, and look at the (3,1) entry: 
\begin{equation}
r_{31} (c_1+c_2\sqrt{p_1}) +(-c_2+c_1\sqrt{p_1})=0.
\end{equation}
Such $r_{31}\in\Q$ does not exist because of $(c_1,c_2)\neq(0,0)$ (see (\ref{eq:Aprime-GammafQ-prm})).
Similarly, we can also see that there is no way this equality holds for the (4,2) entry. 

\vspace{5mm}

The remaining task is to prove that the condition 8(strong) is satisfied in the case (A')--(\ref{eq:Aprime-B+iomega-RCFT}).
That is to construct a Hodge isomorphism $\phi^\ast:H^1(T^4;\Q)\to H^1(T^4_\circ;\Q)$ satisfying $\phi^*|_{\Gamma_{b\Q}^\vee} ={\rm id}_{\Gamma_{b\Q}^\vee}$ for each one of geometric SYZ-mirrors for a set of data $(T^4;G,B;I)$ in the case of (A')--(\ref{eq:Aprime-B+iomega-RCFT}).

To do so, think of a geometric SYZ-mirror for $(\Gamma_f,\Gamma_b)$, and derive a condition for such a Hodge isomorphism $\phi^\ast$, first.
The derivation goes parallel to that for (\ref{eq:Aprime-isogeny-compare}), except for the fact that $E_1$ and $E_2$ are isogenous to $E_1^\circ$ and $E_2^\circ$, respectively. Consequently we obtain
\begin{equation}
\left(\hspace{-5pt}\begin{array}{c|c}
r_{kl} & \hspace{-3pt}\begin{array}{c}0\\\bm{1}\end{array}
\end{array}\hspace{-6pt}\right)
\begin{pmatrix}
c_1+c_2\sqrt{p_1} & \\
& c_3+c_4\sqrt{p_2} \\
-c_2+c_1\sqrt{p_1} & \\
& -c_4+c_3\sqrt{p_2}
\end{pmatrix}
=
\begin{pmatrix}
1 & \\
 & 1\\
 \rho'_1 & \\
 & \rho'_2
\end{pmatrix}
\begin{pmatrix}
\theta'_1 & \\
 & \theta'_2
\end{pmatrix},
\label{eq:Aprime-isogeny-compare-RCFT}
\end{equation}
where $r_{kl}\in\Q$ and $\theta'_i\in\Q(\sqrt{p_i})^\times$ are parameters for $\phi^\ast$, while $c_\text{1--4}\in\Q$ and $\rho'_i\in\Q(\sqrt{p_i})$ depend on $(\Gamma_f,\Gamma_b)$ (see footnote \ref{fn:mirror-dependence-of-rho}).

Now, it is easy to see that a solution $(r_{kl},\theta'_i)$ to (\ref{eq:Aprime-isogeny-compare-RCFT}) exists.
The equation (\ref{eq:Aprime-isogeny-compare-RCFT}) at six out of the $4\times2$ entries imposes 
\begin{align}
&r_{12}=r_{21}=r_{32}=r_{41}=0,\\
&\theta'_1 = r_{11} (c_1+c_2\sqrt{p_1}),\quad \theta'_2 = r_{22} (c_3+c_4\sqrt{p_2}),
\end{align}
and the remaining variables $r_{11},r_{31},r_{22},r_{42}\in\Q$ 
are subject 
\begin{align}
-c_2+c_1\sqrt{p_1}&=\left( \rho'_1 r_{11}-r_{31}\right)(c_1+c_2\sqrt{p_1}),\\
-c_4+c_3\sqrt{p_2}&=\left( \rho'_2 r_{22}-r_{42}\right)(c_3+c_4\sqrt{p_2})
\end{align}
because of (\ref{eq:Aprime-isogeny-compare-RCFT}) at the 
two remaining entries. 
A solution $(r_{11},r_{31},r_{22},r_{42})$ is determined by elementary algebra in the imaginary quadratic fields $\Q(\sqrt{p_1})$ and $\Q(\sqrt{p_2})$.
Moreover, $r_{11}$ and $r_{22}$ turn out to be non-zero, which means that $\phi^\ast$ is invertible.
Since the above argument holds for arbitrary choice of $(\Gamma_f,\Gamma_b)$, this completes the proof of the condition 8(strong) for the case (A')--(\ref{eq:Aprime-B+iomega-RCFT}).

\subsection{The Case (B, C)}
\label{subsec:cond8-caseBC}

Let us show first that the condition 8(weak) is not satisfied 
by a set of data $(T^4;G,B;I)$ in the case 
(B,C)--(\ref{eq:BC-B+iomega};$-$); we do so in a proof by contradiction.
The outline of the proof is the same as in section \ref{subsec:cond8-caseAprm}; 
it just takes a little more time to do so. 

Assume that a set of data $(T^4;G,B;I)$ in the case (B,C)--(\ref{eq:BC-B+iomega};$-$) satisfies the condition 8(weak); 
let $(\Gamma_f,\Gamma_b)$ be the choice for the geometric SYZ-mirror, and $\phi^\ast:H^1(T^4;\Q)$ $\to H^1(T^4_\circ;\Q)$ 
the Hodge isomorphism in the condition. 
We will write down explicitly (i) what is implied by the condition $\phi^*|_{\Gamma_{b\Q}^\vee} ={\rm id}_{\Gamma_{b\Q}^\vee}$ and (ii) what is implied by $\phi^\ast$ being a Hodge isomorphism, and then show that there cannot be such 
$\phi^*$ satisfying both (i) and (ii). 

(i) The condition $\phi^\ast|_{\Gamma_{b\Q}^\vee}=\mathrm{id}_{\Gamma_{b\Q}^\vee}$ can be expressed in the form of 
\begin{align}
\phi^\ast
\begin{pmatrix}
\hat{c}' & \hat{d}' & \hat{e} & \hat{f}
\end{pmatrix}
=
\begin{pmatrix}
c & d & \hat{e} & \hat{f}
\end{pmatrix}
\left(\hspace{-5pt}\begin{array}{c|c}
r_{kl} & \hspace{-3pt}\begin{array}{c}0\\\bm{1}\end{array}
\end{array}\hspace{-6pt}\right),
\label{eq:BC-cond8-asEq}
\end{align}
by using the basis $\{ \hat{c}', \hat{d}', \hat{e}, \hat{f} \}$ of $H^1(T^4;\Q)$ and the basis $\{c,d,\hat{e},\hat{f}\}$ of $H^1(T^4_\circ;\Q)$ explained in (\ref{eq:BC-H^1-mirror}) and (\ref{eq:BC-c_1c_3nonzero-e'f'chat'dhat'}); 
here, $(r_{kl})$ is a $4\times2$ $\Q$\hspace{1pt}-valued matrix. We only deal with the cases with $(c_1,c_3)\neq(0,0)$ here, 
as we did in section \ref{subsec:BC-1-7}; the following discussion can be repeated for the case of $(c_2,c_4)\neq(0,0)$ easily.

Now, the condition (\ref{eq:BC-cond8-asEq}) implies that the holomorphic basis 
(cf. footnote \ref{fn:comment-normalization-dz})
\begin{align}
\begin{pmatrix}
(c_1^2-dc_3^2)dz^1 & (c_1^2-dc_3^2)dz^2
\end{pmatrix}
& =
\begin{pmatrix}
\hat{c}' & \hat{d}' & \hat{e} & \hat{f}
\end{pmatrix}
\begin{pmatrix}
c_1 & c_3 & c_2 & c_4 \\
dc_3 & c_1 & dc_4 & c_2\\
 & & 1 & \\
 & & & 1
\end{pmatrix}
\begin{pmatrix}
1 & 1\\
\tau_{++}(y) & \tau_{-+}(y)\\
\tau_{++}(x) & \tau_{-+}(x)\\
\tau_{++}(xy) & \tau_{-+}(xy)
\end{pmatrix}
\label{eq:BC-isogeny-on-Q-basis-before}
\end{align}
should be mapped by $\phi^\ast$ into
\begin{align}
\phi^\ast_\C
\begin{pmatrix}
(c_1^2-dc_3^2)dz^1 & (c_1^2-dc_3^2)dz^2
\end{pmatrix}
=&
\begin{pmatrix}
c & d & \hat{e} & \hat{f}
\end{pmatrix}
\left(\hspace{-5pt}\begin{array}{c|c}
r_{kl} & \hspace{-3pt}\begin{array}{c}0\\\bm{1}\end{array}
\end{array}\hspace{-6pt}\right)
\begin{pmatrix}
\tau_{++}(\Gamma) & \tau_{-+}(\Gamma) \\
\tau_{++}(\Gamma y) & \tau_{-+}(\Gamma y) \\
\tau_{++}(x) & \tau_{-+}(x) \\
\tau_{++}(xy) & \tau_{-+}(xy)
\end{pmatrix},
\label{eq:BC-isogeny-on-Q-basis}
\end{align}
where $\Gamma:=c_1+c_3y+c_2x+c_4xy\in K$.

(ii) Next, we will translate the condition that $\phi^\ast$ is a Hodge isomorphism.
The images of the holomorphic basis elements 
\begin{align}
\phi^\ast_\C
\begin{pmatrix}
(c_1^2-dc_3^2)dz^1 & (c_1^2-dc_3^2)dz^2
\end{pmatrix}
=:
\begin{pmatrix}
dz^1_\circ & dz^2_\circ
\end{pmatrix}
%
%
\label{eq:BC-isom-diagonal}
%
\end{align}
are eigenvectors of the action of the endomorphism field 
${\rm End}(H^1(T^4_\circ;\Q))^{\rm Hdg}$, and generate the Hodge 
(1,0) component of $H^1(T^4_\circ;\C)$; when the endomorphism fields 
are identified through ${\rm Ad}_{\phi^*}: K \cong {\rm End}(H^1(T^4;\Q))^{\rm Hdg} 
\ni \xi \longmapsto \phi^* \circ \xi \circ (\phi^*)^{-1} \in {\rm End}(H^1(T^4_\circ;\Q))^{\rm Hdg}$, the eigenvalues 
of the endomorphisms on $dz^1_\circ$ and $dz^2_\circ$ are given by the 
embeddings $\tau_{++}$ and $\tau_{-+}$, respectively. 
We note also that both $dz^1_\circ$ and $dz^2_\circ$ are normalized so that they 
return a rational value for some rational elements in $H_1(T^4_\circ;\Q)$.


On the other hand, there is an easy way to describe how an ${\rm End}(H^1(T^4_\circ;\Q))^{\rm Hdg}$-diagonal 
basis $\{ dz^{1'}_\circ, dz^{2'}_\circ\}$ of $H^{1,0}(T^4_\circ;\C)$ is related to the rational 
basis $\{c,d,\hat{e},\hat{f}\}$ of $H^1(T^4_\circ;\Q)$
(as reviewed and then used heavily in \cite{Kidambi:2022wvh}). 
This is done, in principle, by working out the complex structure of the mirror $T^4_\circ$
(cf. (\ref{eq:mirror-cpx-struc})). Instead, we will use an easy way available for 
complex tori to identify the (1,0) component with respect to the vertical 
Hodge structure on $W_v^1/W_v^3$---like we did in (\ref{eq:BC-Wv3-charge1-basis})---and 
translate the information to the mirror side by using the map of D-brane charges (cohomology groups). 
This latter strategy enables us to save time a little more, because we have done the latter calculation 
already in a special case in \cite[\S5.3.1]{Kidambi:2022wvh} (see footnote \ref{fn:BC-charge1-KOW22}).

As a first step, let us identify the subspace $W^1_v \subset H^*(T^4;\Q)$ and introduce 
a rational basis of $W_v^1/W_v^3$. The basis elements 
$\{c,d,\hat{e},\hat{f}\}$ of $H^1(T^4_\circ;\Q)\cong W_{h\circ}^1/W_{h\circ}^3$ on the mirror 
side are identified with the followings by the mirror map of the D-brane charges:
\begin{align}
g^\ast(c) & \in s' \hat{d}'+W_v^3,
\label{eq:BC-pullback-of-W^1-c-rigorous}\\
g^\ast(d) & \in -s'\hat{c}'+W_v^3,
\label{eq:BC-pullback-of-W^1-d-rigorous}\\
g^\ast(\hat{e}) & \in s''\hat{c}'\hat{d}'\hat{e}+W_v^3,
\label{eq:BC-pullback-of-W^1-ehat-rigorous}\\
g^\ast(\hat{f}) & \in s''\hat{c}'\hat{d}'\hat{f}+W_v^3,
\label{eq:BC-pullback-of-W^1-fhat-rigorous}
\end{align}
where $s := [\Gamma_f : (\Z c \oplus \Z d)] \in \Q$, $s' := s/(c_1^2-dc_3^2)$ and 
$s'':= s/(c_1^2-dc_3^2)^2$.

The next step is to identify the Hodge (1,0) component of $(W_v^1/W_v^3)\otimes\C$ in terms of the 
rational basis\footnote{
We use the same notation in this article for both an element of $W_v^1$ and its equivalence class 
in $W_v^1/W_v^3$. 
} %
$\{\hat{c}',\hat{d}',\hat{c}'\hat{d}'\hat{e},\hat{c}'\hat{d}'\hat{f}\}$ of $W_v^1/W_v^3$. 
The (1,0) component is given by $\mho \wedge g^*( H^1(T^4_\circ;\C)/(\Gamma_{b\Q}^\vee\otimes \C) )$, 
which is generated by  
\begin{align}
&\mho\hat{c}'=\hat{c}'+Z_1e_1\hat{c}'+Z_2e_2\hat{c}' \quad \text{and}
\label{eq:BC-mhochat'}\\
&\mho\hat{d}'=\hat{d}'+Z_1e_1\hat{d}'+Z_2e_2\hat{d}' .
\label{eq:BC-mhodhat'}
\end{align}

It is not difficult\footnote{
\label{fn:BC-charge1-KOW22}
When $(c_1,c_2,c_3,c_4)=(1,0,0,0)$, eq.~(\ref{eq:BC-W1-CM}) coincides with
\cite[eq.~(5.38)]{Kidambi:2022wvh}.
} %
to reorganize the basis elements $\{\mho \hat{c}', \mho \hat{d}'\}$ 
to find eigenvectors of the endomorphisms of the vertical Hodge structure on $W^1_v/W^3_v$. 
\begin{align}
& (g^*(dz^{1'}_\circ), g^*(dz^{2'}_\circ))
:= \begin{pmatrix}
\mho\hat{c}' & \mho\hat{d}'
\end{pmatrix}
\begin{pmatrix}
1 & 1\\
\mp\sqrt{d} & \pm\sqrt{d}
\end{pmatrix}\\
&=\begin{pmatrix}
\hat{c}' & \hat{d}' & -e_1\hat{d}' & -e_2\hat{d}'
\end{pmatrix}
\begin{pmatrix}
1 & 1\\
\tau_{++}(\mp y) & \tau_{-+}(\mp y) \\
\tau_{++}(\Xi_\pm) & \tau_{-+}(\Xi_\pm)\\
\tau_{++}(\pm\Xi_\pm y) & \tau_{-+}(\pm\Xi_\pm y)
\end{pmatrix}
\label{eq:BC-W1-CM}\\
&=
\begin{pmatrix}
\hat{c}' & \hat{d}' & \frac{\hat{c}'\hat{d}'\hat{e}}{(c_1^2-dc_3^2)^2} & \frac{\hat{c}'\hat{d}'\hat{f}}{(c_1^2-dc_3^2)^2}
\end{pmatrix}
\begin{pmatrix}
1 & & & \\
& 1 & & \\
& & c_1 & c_3\\
& & dc_3 & c_1
\end{pmatrix}
\begin{pmatrix}
1 & 1\\
\tau_{++}(\mp y) & \tau_{-+}(\mp y) \\
\tau_{++}(\Xi_\pm) & \tau_{-+}(\Xi_\pm)\\
\tau_{++}(\pm\Xi_\pm y) & \tau_{-+}(\pm\Xi_\pm y)
\end{pmatrix},
\label{eq:BC-mhoc'mhod'-c'd'c'd'ec'd'f-CM}
\end{align}
where $\Xi_\pm\in K$ was introduced in (\ref{eq:Xi}); 
it is obvious from this expression (cf.\ \cite[Lemma A.11]{Kidambi:2022wvh})
that they are eigenstates of $K \cong {\rm End}(W^1_v/W^3_v)^{\rm Hdg}$. 
Although we are working over the case (B, C)--(\ref{eq:BC-B+iomega}; $-$) 
at this moment, we will have to repeat the same computation for the case
(B, C)--(\ref{eq:BC-B+iomega}; $+$) later in this section \ref{subsec:cond8-caseBC}. 
So, we have done the computation for both cases above; the sign choice below 
($-$ in $\pm$ and $+$ in $\mp$) is for the case (\ref{eq:BC-B+iomega};$-$) 
and the sign choice above for the case (\ref{eq:BC-B+iomega}; $+$).

The final step is to compute the endomorphism-eigenstates $\{ dz^{1'}_\circ, dz^{2'}_\circ\}$
of $H^1(T^4_\circ;\Q)\otimes \C$. 
Using (\ref{eq:BC-pullback-of-W^1-c-rigorous})--(\ref{eq:BC-pullback-of-W^1-fhat-rigorous}), 
we obtain
\begin{align}
\begin{pmatrix}
dz^{1'}_\circ & dz^{2'}_\circ
\end{pmatrix}
=&
\begin{pmatrix}
-\tilde{s}'d & \tilde{s}'c & \tilde{s}\hat{e} & \tilde{s}\hat{f}
\end{pmatrix}
\begin{pmatrix}
1 & & & \\
& 1 & & \\
& & c_1 & c_3\\
& & dc_3 & c_1
\end{pmatrix}
\begin{pmatrix}
1 & 1\\
\tau_{++}(\mp y) & \tau_{-+}(\mp y) \\
\tau_{++}(\Xi_\pm) & \tau_{-+}(\Xi_\pm)\\
\tau_{++}(\pm\Xi_\pm y) & \tau_{-+}(\pm\Xi_\pm y)
\end{pmatrix}\\
=&
\begin{pmatrix}
c & d & \hat{e} & \hat{f}
\end{pmatrix}
\begin{pmatrix}
\tau_{++}(\mp\tilde{s}'y) & \tau_{-+}(\mp\tilde{s}'y) \\
-\tilde{s}' & -\tilde{s}' \\
\tau_{++}(\tilde{s}(c_1\pm c_3y)\Xi_\pm) & \tau_{-+}(\tilde{s}(c_1\pm c_3y)\Xi_\pm)\\
\tau_{++}(\tilde{s}(dc_3\pm c_1y)\Xi_\pm) & \tau_{-+}(\tilde{s}(dc_3\pm c_1y)\Xi_\pm)
\end{pmatrix},
\label{eq:BC-mirror-hol-basis-vs-rat-basis}\\
&\text{where}\quad \tilde{s}':=s'^{-1}, \quad \tilde{s}:=s^{-1} \in\Q^\times.
\end{align}

The diagonal basis $\{ dz^1_\circ, dz^2_\circ\}$ in (\ref{eq:BC-isom-diagonal})
and another basis $\{ dz^{1'}_\circ, dz^{2'}_\circ\}$ here are not guaranteed 
to be identical, but $dz^1_\circ$ [resp. $dz^2_\circ$] should be a complex multiple of 
$dz^{1'}_\circ$ [resp. $dz^{2'}_\circ$]. Furthermore, the proportionality constant 
should be found within $\tau_{++}(K)$ [resp. $\tau_{-+}(K)$], which follows from 
the fact that both $dz^1_\circ$ and $dz^{1'}_\circ$ return (not necessarily identical) 
rational numbers to (not necessarily identical) rational elements in $H_1(T^4_\circ;\Q)$. 
So, to conclude, the condition (ii) that $\phi^*$ is a Hodge isomorphism is translated to the 
existence of elements $\theta_+, \theta_- \in K^\times$ so that 
\begin{align}
&\phi^\ast_\C
\begin{pmatrix}
(c_1^2-dc_3^2)dz^1 & (c_1^2-dc_3^2)dz^2
\end{pmatrix} \label{eq:BC-isogeny-on-C-basis} \\
&=
\begin{pmatrix}
c & d & \hat{e} & \hat{f}
\end{pmatrix}
\begin{pmatrix}
\tau_{++}(\mp\tilde{s}'y) & \tau_{-+}(\mp\tilde{s}'y) \\
-\tilde{s}' & -\tilde{s}' \\
\tau_{++}(\tilde{s}(c_1\pm c_3y)\Xi_\pm) & \tau_{-+}(\tilde{s}(c_1\pm c_3y)\Xi_\pm)\\
\tau_{++}(\tilde{s}(dc_3\pm c_1y)\Xi_\pm) & \tau_{-+}(\tilde{s}(dc_3\pm c_1y)\Xi_\pm)
\end{pmatrix}
\begin{pmatrix}
\tau_{++}(\theta_+) & \\
 & \tau_{-+}(\theta_-)
\end{pmatrix}.
  \nonumber
\end{align}

\vspace{5mm}

Now that the properties (i) and (ii) in the condition 8 have been paraphrased as (\ref{eq:BC-isogeny-on-Q-basis}) and (\ref{eq:BC-isogeny-on-C-basis}), respectively, let us see that there is no common solution $(r_{kl},\theta_+, \theta_-)$ to (\ref{eq:BC-isogeny-on-Q-basis}) and (\ref{eq:BC-isogeny-on-C-basis};$-$) for any $(c_\text{1--4},  \tilde{s},\tilde{s}')$; this then implies that there is no choice $(\Gamma_f,\Gamma_b)$ for a geometric SYZ-mirror where the condition 8(weak) 
holds true in the case of (B,C)--(\ref{eq:BC-B+iomega};$-$).

Indeed, the compatibility condition on $(r_{k\ell}, \theta_+, \theta_-)$ is 
\begin{align}
&\left(\hspace{-5pt}\begin{array}{c|c}
r_{kl} & \hspace{-3pt}\begin{array}{c}0\\\bm{1}\end{array}
\end{array}\hspace{-6pt}\right)
\begin{pmatrix}
\tau_{++}(\Gamma) & \tau_{-+}(\Gamma) \\
\tau_{++}(\Gamma y) & \tau_{-+}(\Gamma y) \\
\tau_{++}(x) & \tau_{-+}(x) \\
\tau_{++}(xy) & \tau_{-+}(xy)
\end{pmatrix}
\nonumber\\
&=
\begin{pmatrix}
\tau_{++}(\mp\tilde{s}'y) & \tau_{-+}(\mp\tilde{s}'y) \\
-\tilde{s}' & - \tilde{s}'\\
\tau_{++}(\tilde{s}(c_1\pm c_3y)\Xi_\pm) & \tau_{-+}(\tilde{s}(c_1\pm c_3y)\Xi_\pm)\\
\tau_{++}(\tilde{s}(dc_3\pm c_1y)\Xi_\pm) & \tau_{-+}(\tilde{s}(dc_3\pm c_1y)\Xi_\pm)
\end{pmatrix}
\begin{pmatrix}
\tau_{++}(\theta_+) & \\
 & \tau_{-+}(\theta_-)
\end{pmatrix}
\label{eq:BC-isog-C-basis-vs-Q-basis}
\end{align}
when $(c_1,c_3) \neq 0$. The two elements $\theta_+, \theta_- \in K^\times$ have to be identical 
for the relations in the second row to hold. Now, the $4\times2$ relations among algebraic numbers 
in (\ref{eq:BC-isog-C-basis-vs-Q-basis}) are regarded as four relations among the elements in the CM field $K$.
To see that there is no solution $(r_{kl},\theta)$ to (\ref{eq:BC-isog-C-basis-vs-Q-basis};$-$), it suffices to focus on the lower two relations:
\begin{align}
&r_{31}\Gamma+r_{32}\Gamma y+x=\tilde{s}(c_1\pm c_3y)\Xi_\pm\theta,
\label{eq:BC-isog-condition-1}\\
&r_{41}\Gamma+r_{42}\Gamma y+xy=\tilde{s}(dc_3\pm c_1y)\Xi_\pm\theta.
\label{eq:BC-isog-condition-2}
\end{align}
Comparing $(\ref{eq:BC-isog-condition-1})\times(\pm y)$ and (\ref{eq:BC-isog-condition-2}), we can eliminate $\theta$ and obtain
\begin{align}
[\pm dr_{32}-r_{41}+(\pm r_{31}-r_{42})y]\Gamma=&\ axy, 
\label{eq:BC-isog-condition-final}\\
&\ a=\left\{\begin{array}{ll}
0 & \text{in the case (B,C)--(\ref{eq:BC-B+iomega};$+$)}, \\
2 & \text{in the case (B,C)--(\ref{eq:BC-B+iomega};$-$)}.
\end{array}\right.
\end{align}
Recalling that $\Gamma=c_1+c_3y+c_2x+c_4xy$ and $(c_1,c_3)\neq(0,0)$, some easy algebra in the number field $K$ shows that this equation (\ref{eq:BC-isog-condition-final};$a=2$) for the case (B,C)--(\ref{eq:BC-B+iomega};$-$) does not have a solution $(r_{31},r_{32},r_{41},r_{42})$.

\vspace{5mm}

The remaining task is to prove that the condition 8(strong) is satisfied in the case (B, C)--(\ref{eq:BC-B+iomega};$+$).
That is to construct a Hodge isomorphism $\phi^\ast:H^1(T^4;\Q)\to H^1(T^4_\circ;\Q)$ satisfying $\phi^*|_{\Gamma_{b\Q}^\vee} ={\rm id}_{\Gamma_{b\Q}^\vee}$ for each one of geometric SYZ-mirrors for a set of data $(T^4;G,B;I)$ in the case 
(B, C)--(\ref{eq:BC-B+iomega};$+$).

To do so, we will find a solution $(r_{kl},\theta)$ to the equation (\ref{eq:BC-isog-C-basis-vs-Q-basis};$+$) for a given $(c_\text{1--4},\tilde{s},\tilde{s}')$.
Just as a reminder, $(r_{kl},\theta)$ are parameters for $\phi^\ast$, whereas $(c_\text{1--4},\tilde{s},\tilde{s}')$ are determined by a given $(\Gamma_f,\Gamma_b)$ for a geometric SYZ-mirror.

In fact, we can solve the equation (\ref{eq:BC-isog-C-basis-vs-Q-basis};$+$) as follows.
As already discussed above, the lower two rows of the equation (\ref{eq:BC-isog-C-basis-vs-Q-basis};$+$) lead to equations (\ref{eq:BC-isog-condition-1};$+$), (\ref{eq:BC-isog-condition-2};$+$), and (\ref{eq:BC-isog-condition-final}; $a=0$).
The last one (\ref{eq:BC-isog-condition-final}; $a=0$) is equivalent to
\begin{align}
r_{42}=r_{31},\quad r_{41}=dr_{32}.
\label{eq:BC-isog-solution-1}
\end{align}
In the same way, from the upper two rows of the equation (\ref{eq:BC-isog-C-basis-vs-Q-basis};$+$), we find
\begin{align}
  r_{11} = d r_{22},\quad  r_{12} = r_{21},
\label{eq:BC-isog-solution-2}
\end{align}
and
\begin{align}
&r_{21}\Gamma+r_{22}\Gamma y= - \tilde{s}'\theta.
\label{eq:BC-isog-condition-3}
\end{align}
Now, by eliminating $\theta$ from (\ref{eq:BC-isog-condition-1};$+$) and (\ref{eq:BC-isog-condition-3}), we have
\begin{align}
\tilde{s}'r_{31} + \tilde{s}'r_{32}y + r_{21}\Xi'_+ + r_{22}\Xi'_+y=-\tilde{s}'x\Gamma^{-1},
\label{eq:BC-isog-condition-4}
\end{align}
where $\Xi'_+:=\tilde{s}(c_1+c_3y)\Xi_+$.
Since $1,y,\Xi'_+,\Xi'_+y\in K$ are linearly independent over $\Q$, this equation (\ref{eq:BC-isog-condition-4}) determines 
$(r_{31},r_{32},r_{21},r_{22})$ uniquely.

It is easy to confirm that the equation (\ref{eq:BC-isog-C-basis-vs-Q-basis};$+$) is satisfied whenever the parameters $(r_{kl},\theta)$ 
are subject to the relations:
\begin{align*}
&(\ref{eq:BC-isog-condition-4}): \tilde{s}'r_{31}+\tilde{s}'r_{32}y + r_{21}\Xi'_+ + r_{22}\Xi'_+y=-\tilde{s}'x\Gamma^{-1},\\
&(\ref{eq:BC-isog-solution-1}): r_{42}=r_{31},\quad r_{41}=dr_{32},\\
&(\ref{eq:BC-isog-solution-2}):  r_{11}= d r_{22},\quad r_{12} = r_{21},\\
&(\ref{eq:BC-isog-condition-3}): \tilde{s}'\theta= -(r_{21}\Gamma+r_{22}\Gamma y).
\end{align*}
So, indeed, there is a solution $(r_{k\ell}, \theta)$ common to both (i) and (ii). Furthermore, the Hodge morphism $\phi^\ast$ specified by this $(r_{kl},\theta)$ 
is an isomorphism; if $\theta$ were zero (and hence $\phi^\ast$ were not invertible), then $\Gamma$, $\Gamma y$, and $x$ would not be linear independent over $\Q$, 
according to (\ref{eq:BC-isog-condition-1}).
Since the above argument holds for arbitrary choice of a geometric SYZ-mirror, this completes the proof of the condition 8(strong) for the case (B,C)--(\ref{eq:BC-B+iomega};$+$).

\subsection{The Case (A)}
\label{subsec:cond8-caseA}

We have already dealt with the cases (A') and (B, C) earlier in this section, but there is one more case we should consider: the case (A). As reviewed briefly at the beginning of section \ref{sec:relations}, a set of data $(T^4;G,B;I)$ satisfying 1--3(a) is classified into one of the four cases (A), (A'), (B and C) based on its complex structure $(T^4;I)$, and such a set of data in the case (A) always gives rise to a rational $\cN=(1,1)$ SCFT. The case (A) is when the abelian surface $(T^4,I)$ 
is isogenous to the product $E\times E$ of CM elliptic curves (cf \cite[Lemma 2.22]{Kidambi:2022wvh}). 

Therefore, to affirm that the set of conditions 1, 2, 3(a) and 8 (either weak or strong) on a set of data $(T^4;G,B;I)$ is necessary and sufficient for the corresponding $\cN=(1,1)$ SCFT to be rational, we also have to prove that the condition 8(strong) is always satisfied in the case (A). 
That is to show the existence of a Hodge isomorphism $\phi^\ast:H^1(T^4;\Q)\to H^1(T^4_\circ;\Q)$ satisfying $\phi^\ast|_{\Gamma_{b\Q}^\vee}=\id_{\Gamma_{b\Q}^\vee}$ for any choice of $(\Gamma_f,\Gamma_b)$ for a geometric SYZ-mirror in the case (A). 

To get started, let us have a few words about rational and holomorphic bases of $H^1(T^4)$ and 
$H^1(T^4_\circ)$ to be used in the analysis. Let $\{c,d\}$ be a basis of $\Gamma_{f\Q}$, 
and $\{e,f\}$ that of $\Gamma_{b\Q}$;
the subspaces $\Gamma_{f\Q}^\vee$ and $\Gamma_{b\Q}^\vee$ of $H^1(T^4_\circ;\Q)$ are generated by 
$\{\hat{c},\hat{d}\}$ and $\{ \hat{e}, \hat{f}\}$, respectively; here, 
$\{\hat{c},\hat{d},\hat{e},\hat{f}\}$ is the dual basis of $\{ c,d,e,f\}$.  

An easiest basis of $H^{1,0}(T^4;\C)$ over $\C$ one thinks of is obtained by choosing 
a basis of $H^{1,0}(E;\C)$ of the two CM elliptic curves, $dz^1$ and $dz^2$, and pull them 
back to $H^{1,0}(T^4;\C)$. When $dz^1$ and $dz^2$ are normalized so that they return 
rational values to a rational cycle of the two $E$'s, their pullbacks on $T^4$---denoted 
by the same $dz^1$ and $dz^2$---are related to the rational 
basis by 
\begin{align}
\begin{pmatrix}
dz^1 & dz^2
\end{pmatrix}
=
\begin{pmatrix}
\hat{c} & \hat{d} & \hat{e} & \hat{f}
\end{pmatrix}
\begin{pmatrix}
\lambda_{11} & \lambda_{12}\\
\vdots & \vdots\\
\lambda_{41} & \lambda_{42}
\end{pmatrix},
\label{eq:A-hol-vs-rat-basis}
\end{align}
with the coefficients $\lambda_{ij}$ in the imaginary quadratic field $\Q(\sqrt{p})$ 
of the complex multiplication of $E$.

We may change the holomorphic basis and rational basis a little bit so that the 
analysis later in this section \ref{subsec:cond8-caseA} is easier. 
Note first that the upper $2 \times 2$ block of the coefficient matrix in (\ref{eq:A-hol-vs-rat-basis})
is invertible. This is because the holomorphic $n$-form on the $T^n$ fiber should have 
a non-zero period in the SYZ mirror correspondence. So, one may change the basis $\{ dz^1, dz^2\}$
by a ${\rm GL}_2(\Q(\sqrt{p}))$ transform so that 
the upper $2\times 2$ block in (\ref{eq:A-hol-vs-rat-basis})
is the identity matrix. Furthermore, it is possible to rearrange the rational basis of $\Gamma_{b\Q}^\vee$, 
denoted by $\{ \hat{e}', \hat{f}'\}$ now, so that the coefficient $\lambda_{31}$ becomes rational. To summarize, 
there are a rational basis $\{ \hat{c}, \hat{d}\}$ of $\Gamma_{f\Q}^\vee$, $\{ \hat{e}', \hat{f}'\}$ 
of $\Gamma_{b\Q}^\vee$, and a holomorphic basis $\{ d\tilde{z}^1, d\tilde{z}^2\}$ of $H^{1,0}(T^4;\C)$
so that 
\begin{align}
\begin{pmatrix}
d\tilde{z}^1 & d\tilde{z}^2
\end{pmatrix}
=
\begin{pmatrix}
\hat{c} & \hat{d} & \hat{e}' & \hat{f}'
\end{pmatrix}
\begin{pmatrix}
1 & 0\\
0 & 1\\
\widetilde{\lambda}_{31}' & \widetilde{\lambda}_{32}'\\
\widetilde{\lambda}_{41}' & \widetilde{\lambda}_{42}'
\end{pmatrix}, \quad \text{where} \quad \widetilde{\lambda}_{31}'\in\Q,
\label{eq:A-retake-rat}
\end{align}
and $\widetilde{\lambda}_{32}',\widetilde{\lambda}_{41}',\widetilde{\lambda}_{42}'\in\Q(\sqrt{p})$.

On the mirror side, we can regard  $\{c,d,\hat{e}',\hat{f}'\}$ as a rational basis of $H^1(T^4_\circ;\Q)$ (see (\ref{eq:Aprime-H^1-mirror}) and footnote \ref{fn:construction-of-mirror}).
Since a set of data $(T^4;G,B;I)$ satisfying 1--3(a) in the case (A) always gives rise to a rational $\cN=(1,1)$ SCFT, we can freely use the properties 1--7; in particular, $T^4_{I\circ}$ is isogenous to $T^4_I$ 
(see \cite[Prop.\ 3.10]{Chen:2005gm}), and hence to $E\times E$. Repeating the same argument as 
in the case of $T^4$, one can find a basis $\{ d\tilde{z}^1_\circ, d\tilde{z}^2_\circ \}$ of $H^{1,0}(T^4_\circ;\C)$
so that 
\begin{align}
\begin{pmatrix}
d\tilde{z}_\circ^1 & d\tilde{z}_\circ^2
\end{pmatrix}
=
\begin{pmatrix}
c & d & \hat{e}' & \hat{f}'
\end{pmatrix}
\begin{pmatrix}
1 & 0\\
0 & 1\\
\widetilde{\rho}'_{31} & \widetilde{\rho}'_{32}\\
\widetilde{\rho}'_{41} & \widetilde{\rho}'_{42}
\end{pmatrix}
\end{align}
for some coefficients $\widetilde{\rho}'_{ij}$ in $\Q(\sqrt{p})$.

Having done a preparation, we start showing that the condition 8 (strong) is satisfied. 
As we have done earlier, we will translate (i) the condition (\ref{eq:cond-8}) 
and (ii) the Hodge-ness condition of the isomorphism $\phi^*: H^1(T^4;\Q) \rightarrow H^1(T^4_\circ;\Q)$, 
and then find a common solution.  

(i) The condition $\phi^\ast|_{\Gamma_{b\Q}^\vee}=\id_{\Gamma_{b\Q}^\vee}$ is equivalent to the existence of a $4\times2$ $\Q$\hspace{1pt}-valued matrix $(r_{kl})$ such that
\begin{align}
\phi^\ast_\C
\begin{pmatrix}
d\tilde{z}^1 & d\tilde{z}^2
\end{pmatrix}
=&
\begin{pmatrix}
c & d & \hat{e}' & \hat{f}'
\end{pmatrix}
\left(\hspace{-5pt}\begin{array}{c|c}
r_{kl} & \hspace{-3pt}\begin{array}{c}0\\\bm{1}\end{array}
\end{array}\hspace{-6pt}\right)
\begin{pmatrix}
1 & 0\\
0 & 1\\
\widetilde{\lambda}_{31}' & \widetilde{\lambda}_{32}'\\
\widetilde{\lambda}_{41}' & \widetilde{\lambda}_{42}'
\end{pmatrix}.
\label{eq:A-isogeny-on-Q-basis}
\end{align}

(ii) We will write down the condition implied by $\phi^\ast$ being a Hodge 
isomorphism.
In the case (A), we have $\End(H^1(T^4_I;\Q))^\Hdg\cong M_2(\Q(\sqrt{p}))$, 
where $M_2(\Q(\sqrt{p}))$ is the $2\times2$ matrix algebra with the matrix 
entries in $\Q(\sqrt{p})$.
Moreover, all the Hodge isomorphisms $H^1(T^4;\Q)\to H^1(T^4_\circ;\Q)$ form a set that is one to one\footnote{
Recall that $T^4_{I\circ}$ is always isogenous to $T^4_I$ in the case (A).
Let $\psi:T^4_I\to T^4_{I\circ}$ be an isogeny.
The one-to-one correspondence in the main text assigns $\phi\circ\psi^\ast\in[\End(H^1(T^4_\circ;\Q))^\Hdg]^\times$ to a Hodge isomorphism $\phi:H^1(T^4;\Q)\to H^1(T^4_\circ;\Q)$.
} with $[\End(H^1(T^4_\circ;\Q))^\Hdg]^\times\cong M_2(\Q(\sqrt{p}))^\times$.
Therefore, the condition that $\phi^\ast$ is a Hodge isomorphism is equivalent 
to the existence of $(\theta_{ij})\in M_2(\Q(\sqrt{p}))^\times$ such that
\begin{align}
\phi^\ast_\C
\begin{pmatrix}
d\tilde{z}^1 & d\tilde{z}^2
\end{pmatrix}
=&
\begin{pmatrix}
c & d & \hat{e}' & \hat{f}'
\end{pmatrix}
\begin{pmatrix}
1 & 0\\
0 & 1\\
\widetilde{\rho}'_{31} & \widetilde{\rho}'_{32}\\
\widetilde{\rho}'_{41} & \widetilde{\rho}'_{42}
\end{pmatrix}
\begin{pmatrix}
\theta_{11} & \theta_{12}\\
\theta_{21} & \theta_{22}
\end{pmatrix}.
\label{eq:A-isogeny-on-C-basis}
\end{align}

\vspace{5mm}

Now that the properties (i) and (ii) in the condition 8 have been paraphrased as (\ref{eq:A-isogeny-on-Q-basis}) and (\ref{eq:A-isogeny-on-C-basis}), respectively, let us see that there exists a common solution $(r_{kl},\theta_{ij})$ to (\ref{eq:A-isogeny-on-Q-basis}) and (\ref{eq:A-isogeny-on-C-basis}) for a given $(\widetilde{\lambda}'_{ij},\widetilde{\rho}'_{ij})$.
Just as a reminder, $(r_{kl},\theta_{ij})$ are parameters of $\phi^\ast$, whereas $(\widetilde{\lambda}'_{ij},\widetilde{\rho}'_{ij})$ depend on a given $(\Gamma_f,\Gamma_b)$ for a geometric SYZ-mirror.

The compatibility condition of (\ref{eq:A-isogeny-on-Q-basis}) 
and (\ref{eq:A-isogeny-on-C-basis}) is 
\begin{align}
\left(\hspace{-5pt}\begin{array}{c|c}
r_{kl} & \hspace{-3pt}\begin{array}{c}0\\\bm{1}\end{array}
\end{array}\hspace{-6pt}\right)
\begin{pmatrix}
1 & 0\\
0 & 1\\
\widetilde{\lambda}'_{31} & \widetilde{\lambda}'_{32}\\
\widetilde{\lambda}'_{41} & \widetilde{\lambda}'_{42}
\end{pmatrix}
=
\begin{pmatrix}
1 & 0\\
0 & 1\\
\widetilde{\rho}'_{31} & \widetilde{\rho}'_{32}\\
\widetilde{\rho}'_{41} & \widetilde{\rho}'_{42}
\end{pmatrix}
\begin{pmatrix}
\theta_{11} & \theta_{12}\\
\theta_{21} & \theta_{22}
\end{pmatrix}.
\label{eq:A-isog-compare}
\end{align}
This equation (\ref{eq:A-isog-compare}) is read as the following 
relations in the $4\times 2$ matrix entries:
\begin{align}
& \theta_{ij}=r_{ij} \qquad \qquad (i,j = 1,2),
\label{eq:A-solution-0}\\
& r_{31}+\widetilde{\lambda}'_{31}=\widetilde{\rho}'_{31}r_{11}+\widetilde{\rho}'_{32}r_{21},
\label{eq:A-solution-1}\\
& r_{41}+\widetilde{\lambda}'_{41}=\widetilde{\rho}'_{41}r_{11}+\widetilde{\rho}'_{42}r_{21},
\label{eq:A-solution-2}\\
& r_{32}+\widetilde{\lambda}'_{32}=\widetilde{\rho}'_{31}r_{12}+\widetilde{\rho}'_{32}r_{22},
\label{eq:A-solution-3}\\
& r_{42}+\widetilde{\lambda}'_{42}=\widetilde{\rho}'_{41}r_{12}+\widetilde{\rho}'_{42}r_{22}.
\label{eq:A-solution-4}
\end{align}

The parameters $\theta_{ij}$ are fixed relatively to $r_{k\ell} \in \Q$.
The parameters $(r_{11},r_{21},r_{31},r_{41})$ are determined from the 
equations (\ref{eq:A-solution-1}) and (\ref{eq:A-solution-2}) as follows.
From the imaginary parts of (\ref{eq:A-solution-1}) 
and (\ref{eq:A-solution-2}), we find
\begin{align}
\left\{\begin{array}{l}
\widetilde{\rho}'^{(2)}_{31}r_{11}+\widetilde{\rho}'^{(2)}_{32}r_{21}=\widetilde{\lambda}'^{(2)}_{31}(=0),\\
\widetilde{\rho}'^{(2)}_{41}r_{11}+\widetilde{\rho}'^{(2)}_{42}r_{21}=\widetilde{\lambda}'^{(2)}_{41},
\end{array}\right.
\end{align}
where we introduced a notation 
$\mu=\mu^{(1)}+\sqrt{p}\mu^{(2)}$ ($\mu^{(1)},\mu^{(2)}\in\Q$) for 
$\mu\in\Q(\sqrt{p})$.
This system of equations has a unique solution $(r_{11},r_{21})$, because $\begin{pmatrix}\widetilde{\rho}'^{(2)}_{31}&\widetilde{\rho}'^{(2)}_{32}\end{pmatrix}$ and $\begin{pmatrix}\widetilde{\rho}'^{(2)}_{41}&\widetilde{\rho}'^{(2)}_{42}\end{pmatrix}$ are linearly independent; if not, it contradicts the fact that the $4\times4$ matrix $\begin{pmatrix}\bm{1}&\bm{1}\\\widetilde{\rho}'_{ij}&\widetilde{\rho}'^{\mathrm{c.c.}}_{ij}\end{pmatrix}$ is a change-of-basis matrix from the rational one to the complex one of $H^1(T^4_\circ;\C)$, and hence invertible.
Now that we have obtained $(r_{11},r_{21})$, we can also determine the remaining $r_{31}$ and $r_{41}$ from (\ref{eq:A-solution-1}) and (\ref{eq:A-solution-2}), respectively. The parameters $(r_{12},r_{22},r_{32},r_{42})$ are also determined 
in the same way from (\ref{eq:A-solution-3}) and (\ref{eq:A-solution-4}).

This solution $(r_{kl},\theta_{ij})$ specifies a Hodge morphism 
$\phi^\ast:H^1(T^4;\Q)\to H^1(T^4_\circ;\Q)$ satisfying 
$\phi^\ast|_{\Gamma_{b\Q}^\vee}=\id_{\Gamma_{b\Q}^\vee}$.
To make sure that this $\phi^\ast$ is an isomorphism, we have to show 
that $\begin{pmatrix}\theta_{11}&\theta_{12}\\\theta_{21}&\theta_{22}\end{pmatrix}$ or equivalently $\begin{pmatrix}r_{11}&r_{12}\\r_{21}&r_{22}\end{pmatrix}$ 
for this solution is invertible.
We will do this by contradiction as follows.

Assume that the $2\times2$ $\Q$\hspace{1pt}-valued matrix $\begin{pmatrix}r_{11}&r_{12}\\r_{21}&r_{22}\end{pmatrix}$ is not invertible.
This implies that the right-hand sides of the equations (\ref{eq:A-solution-1}) and (\ref{eq:A-solution-3}) coincide up to multiplication by rational constant, and therefore we find that $\widetilde{\lambda}'_{32}$ can be written as a $\Q$-linear combination of $r_{31}$, $r_{32}$ and $\widetilde{\lambda}'_{31}$.
Since $\widetilde{\lambda}'_{31}$ is a rational number as in (\ref{eq:A-retake-rat}), both $\widetilde{\lambda}'_{31}$ and $\widetilde{\lambda}'_{32}$ are real valued, in particular.
This contradicts the fact that the $4\times4$ matrix $\begin{pmatrix}\bm{1}&\bm{1}\\\widetilde{\lambda}'_{ij}&\widetilde{\lambda}'^{\mathrm{c.c.}}_{ij}\end{pmatrix}$ is invertible.
This ends the proof of $\phi^\ast$ being an isomorphism.

Since the above argument holds for arbitrary choice of a geometric SYZ-mirror, this completes the proof of the condition 8(strong) for the case (A).

\section{Towards Complete Characterization of Rational SCFTs}
\label{sec:towards}

As is evident from what we wrote in Introduction, 
this section is still about 2d SCFTs that are interpreted 
as non-linear sigma models with the target spaces and Ricci-flat K\"{a}hler metrics.
The section title has been trimmed down to fit within a single line. 

Let us first write down an updated version of Thm.\ 5.8 
of \cite{Kidambi:2022wvh}, verified for the case of $M=T^4$, 
in a language applicable to any self-mirror family of Ricci-flat 
K\"{a}hler manifolds $M$. 

\vspace{3mm}
{\bf Theorem} (for $M=T^4$) / {\bf Conjecture 1} (for self-mirror $M$): 

Let $M$ be a real $2n$-dimensional manifold which admits a Ricci-flat 
K\"{a}hler metric $G$, and $B$ a closed 2-form on $M$. 
Suppose, further, that the family of such $(M;G,B)$ is self-mirror 
in that $h^{p,q}(M) = h^{n-p,q}(M)$. The non-linear sigma 
model $\cN=(1,1)$ SCFT associated with the data $(M; G, B)$ is 
a rational SCFT if and only if the following conditions are satisfied: 

\begin{enumerate}
\item there exists a polarizable complex structure $I$ with which 
the metric $G$ is compatible and $(M; G; I)$ becomes K\"{a}hler, and 
$B^{(2,0)}=0$; 
\item the complexified K\"{a}hler parameter $(B+i\omega)$, where 
 $\omega(-,-) = 2^{-1}G(I-,-)$ is the K\"{a}hler form, is in the 
algebraic part  $(H^2(M;\Q) \cap H^{1,1}(M;\R))\otimes \C$; 
\item there exists a geometric SYZ-mirror of the $\cN=(1,1)$ SCFT;
  there may be more than one; the data of such a mirror is denoted by $(W;G^\circ, B^\circ; I^\circ)$;
\item the rational Hodge structure on $H^*(M;\Q)$ is of CM-type; 
\item (strong) for {\it any one} of the geometric SYZ-mirror SCFTs, there is 
a Hodge isomoprhism $\phi^*: H^*(M;\Q) \rightarrow H^*(W;\Q)$ such that 
$\phi^*$ is the identity map on the vector subspaces 
$\pi_M^*: H^*(B;\Q) \hookrightarrow H^*(M;\Q)$ and 
$\pi_W^*: H^*(B;\Q) \hookrightarrow H^*(W;\Q)$; here, 
$\pi_M: M \rightarrow B$ and $\pi_W: W \rightarrow B$ are the 
SYZ $T^n$-fibrations over a common base manifold $B$ of real dimension $n$. 
\setcounter{enumi}{4}
\item (weak) 
there {\it exists} a geometric SYZ-mirror SCFT for which 
there is a Hodge isomoprhism $\phi^*: H^*(M;\Q) \rightarrow H^*(W;\Q)$ 
such that $\phi^*$ is the identity map on the vector subspaces 
$\pi_M^*: H^*(B;\Q) \hookrightarrow H^*(M;\Q)$ and 
$\pi_W^*: H^*(B;\Q) \hookrightarrow H^*(W;\Q)$; here, 
$\pi_M: M \rightarrow B$ and $\pi_W: W \rightarrow B$ are the 
SYZ $T^n$-fibrations over a common base manifold $B$ of real dimension $n$. 
$\bullet$
\end{enumerate}
\vspace{3mm}

The set of conditions 1--8(strong/weak) in the earlier sections is equivalent to 
the set of conditions 1--5(strong/weak) here (Thm.\ 1) in the case $M=T^4$ 
(as we argue shortly). In the case of a more general self-mirror $M$, the set of 
conditions 1--5 here (Conj.\ 1) is meant as a proposal/guess for how to generalize. 

Let us first confirm in the case $M=T^4$ that the conditions 1--5 here are equivalent 
to the conditions 1--8 earlier (so that Thm.\ 1 holds for $M=T^4$). 
The conditions 2--6 of Thm.\ 5.8 of \cite{Kidambi:2022wvh} are captured by 
the conditions 2--4 and the existence of $\phi^*$ in the condition 5 here. 
It may appear that the condition 5 here demands more properties on $\phi^*$
than the condition 8 in section \ref{sec:one-more} does;  
$\phi^*$ in Thm.\ 1 is required to be defined on the whole $H^*(M;\Q)$ here 
than on $H^1(M;\Q)$, first of all, and to be the identity on the whole $H^*(B;\Q)$, 
not just on $\Gamma_{b\Q}^\vee \subset H^1(B;\Q) \subset H^*(B;\Q)$, secondly.  
In fact, a Hodge isomorphism on $H^1(M;\Q)_{M=T^4}$ as in the condition 8 
can be extended by the wedge product (cup product) to a Hodge isomorphism on 
$H^*(M;\Q)_{M=T^4}$; the property that $\phi^*$ is identity on the subspace 
$H^*(B;\Q)_{B=T^2}$ also follows automatically from that for $H^1(B;\Q)_{B=T^2}$.
 
The conditions are reorganized as above for the following reason. 
We have found in this article that the condition 8 in 
section \ref{sec:one-more} involves the SYZ torus fibration.
Now that the SYZ torus fibration plays an essential role, it is more natural to use the horizontal Hodge structure on the mirror $H^\ast(W;\Q)$ rather than the vertical Hodge structure on $H^\ast(M;\Q)$.
This is also why the conditions above are stated without referring\footnote{
As a part of necessary conditions, the CM-ness and the Hodge isomorphism 
of the vertical Hodge structure on $T^v_M \otimes \Q$ is still useful. 
} %
to $T^v_M \otimes \Q \subset A(M)\otimes \Q$ as in Thm.\ 5.8 of \cite{Kidambi:2022wvh}.

\vspace{8mm}

Conjecture 1 in the case of a self-mirror $M$ is still stated in the language 
of a classical geometry, such as $(M; G, B; I)$. That is not particularly 
an issue when $M \sim T^{2n}$ and $M \sim {\rm K3}$; for a more general 
self-mirror family, however, treated data and conditions on them should be 
not in terms of classical geometry, but in terms of ${\cal N}=(1,1)$ SCFT. 
A set of data $(M; G, B)$ corresponds to a point (one SCFT) in the moduli 
space of ${\cal N}=(1,1)$ SCFTs of the target space $M$, and the homology 
groups of $H_*(M;\Z)$ to D-brane charges in the ${\cal N}=(1,1)$ SCFT. 
A choice of a complex structure $I$ must be encoded in a choice of 
one weight-1 operator in the left moving sector and one weight-1 
operator in the right moving sector that enhances the ${\cal N}=1$ superconformal 
algebra in each sector to an ${\cal N}=2$ superconformal algebra; this choice 
determines spectral flow operators in the ${\cal N}=(1,1)$ SCFT, and supersymmetry 
charges of the effective field theory after the compactification on ``$(M;G,B)$.''
The Hodge structures on $H^*(M;\Z)$ appearing in the conditions of Conjecture 1 
should be phrased in terms of the central charges of the effective field theory. 
The pull back $\pi^*_M: H^*(B;\Q) \longrightarrow H^*(M;\Q)$ of an SYZ torus fibration 
is captured by introducing a filtration in the space of D-brane charges. 
That will be an outline; details and gaps between the lines should still be 
filled here and there in writing down Conjecture 1 in the abstract language 
of 2d SCFT.  

\vspace{8mm}

Conjecture 1 is certainly a natural extrapolation from the case $M=T^4$
to the cases of $M$'s that forms a self-mirror family. It makes sense, and 
there is no obvious evidence that it is wrong at this moment, at least in the eyes 
of the present authors. What if we think of a more general class of 
families of Ricci-flat K\"{a}hler manifolds $M$ then?

When a family of Ricci-flat K\"{a}hler manifolds $M$ is not self-mirror, 
the conditions 1--3 in Conj.\ 1 still make sense.\footnote{
When we think of $M$ with $h^{2,0}(M)=0$ (for example, 
when $M$ is a Calabi--Yau $n$-fold with $n>2$), the 
conditions 1 and 2 in Conj. 1 are automatically satisfied. 
} %
The condition 5 should be modified, however, because examples of 
rational SCFTs are known for non-self-mirror $M$'s 
(e.g., Gepner constructions), and yet a Hodge isomorphism $\phi^*$ 
in the condition 5 would imply that $b_k(M) = b_k(W)$ for $k=0,1,\cdots, 2n$. 
Relying on a random guess, we propose to modify the conditions as follows:

\vspace{3mm}
{\bf Conjecture} 2 (general): 
Let $M$ be a real $2n$-dimensional manifold which admits a Ricci-flat 
K\"{a}hler metric $G$, and $B$ a closed 2-form on $M$. The non-linear sigma 
model $\cN=(1,1)$ SCFT associated with the data $(M; G, B)$ is 
conjectured to be a rational SCFT if and only if the following conditions 
are satisfied: 

The conditions 1--3 remain the same as in the case of self-mirror $M$'s. 
\begin{enumerate}
\setcounter{enumi}{3}
\item (gen./strong) for any one of the SYZ mirrors, one can find a rational 
Hodge substructure $V_M^{(k)} \subset H^k(M;\Q)$ and also a 
rational Hodge substructure $V_W^{(k)} \subset H^k(W;\Q)$ for each $0 \leq k \leq n$ 
that satisfy the following conditions (a--c); let $\Pi_M^{(k)}$ and $\Pi_W^{(k)}$ 
be the projection from $H^k(M;\Q)$ to $V_M^{(k)}$ and $H^k(W;\Q)$ to $V_W^{(k)}$, respectively; 
first, $\Pi_M^* \circ \pi_M^*: H^k(B;\Q) \rightarrow V_M^{(k)}$ and 
$\Pi_W^{(k)} \circ \pi_W^*: H^k(B;\Q) \rightarrow V_W^{(k)}$ should be injective (a);
furthermore, 
\begin{itemize}
\item [(b)] the rational Hodge structure on $V_M^{(k)}$ is of CM-type, 
\item [(c)] there exists a Hodge morphism $\phi^*: V_M^{(k)} \rightarrow V_W^{(k)}$
  such that $\phi^* \circ (\Pi_M^{(k)} \circ \pi_M^*) = \Pi_W^{(k)} \circ \pi_W^*$. 
\end{itemize}
$\bullet$
\end{enumerate}

\vspace{3mm}

The conditions written above are only meant to be trial versions. 
 One may think of changing the condition 4 from the one for {\it arbitrary} 
 geometric SYZ mirrors as above, to the one for {\it some} geometric SYZ mirrors, 
 when the condition is referred to as the condition 4(gen./weak). 
Besides this strong vs weak variation, there is still a wide variety in modifying 
the conditions; one might demand that the entire $H^k(M;\Q)$ is of CM-type instead 
of the condition 4(b) in Conj.\ 2, or demand a Hodge morphism $\phi^*: H^k(M;\Q) \rightarrow H^k(W;\Q)$
such that $\phi^* \circ \pi_M^* = \pi_W^*$ on $H^k(B;\Q)$ instead 
of the condition 4(c) in Conj.\ 2.  At this moment, there is not much evidence 
to pin down the right characterization conditions from study of SCFTs. 

One may still run a few tests on formal aspects of Conjecture 2 to access 
its credibility. We do so in the rest of this article. 
The condition 4 in Conj.\ 2 as it stands passes the two tests below; 
reference to a substructure $V_M^{(k)}$ is motivated in that context. 

Let us take a moment before jumping into the first test, to prepare 
notations and cultivate intuitions on what the substructure $V_M^{(k)}$
should be like. Suppose that 
\begin{align}
 H^k(M;\Q) \cong \oplus_{\alpha \in {\cal A}^k(M)} \oplus_{a \in \alpha} V_a^{(k)}
    =: \oplus_{\alpha \in {\cal A}^k(M)} V_\alpha^{(k)}
 \end{align}
is a decomposition into simple rational Hodge substructures;\footnote{
There exists a simple substructure decomposition, because the rational 
Hodge structure is polarizable (the condition 1). 
} %
simple rational Hodge substructures labeled by $a \in A^k(M)$ are grouped 
together by Hodge isomorphisms among them;\footnote{
\label{fn:Fermat}
Take the Fermat quintic Calabi--Yau threefold $(M; I)$ as an example (it 
extracts the complex structure information of the Gepner construction 
$(3^{\otimes 5})/\Z_5$), and focus on $H^3(M;\Q)$. There are two 
Hodge-isomorphism classes (i.e., $\#[{\cal A}^3(M)]=2$), denoted by 
$\alpha_3$ and $\alpha_1$. The level-3 component $V_{\alpha_3}^{(3)}$ 
is of 4-dimensions over $\Q$, which contains the Hodge (3,0) and (0,3) components. 
The level-1 component $V_{\alpha_1}^{(3)}$ consists of 50 copies of simple 
rational Hodge substructures, each of which is of 4-dimensions over $\Q$
\cite{shioda1982geometry}.  The authors are curious, in an SYZ $T^3$-fibration of 
a Fermat quintic $\pi_M: M \rightarrow B = S^3$, whether the Poincare dual of the $T^3$ fiber
(which is the pull-back image of the generator of $H^3(B;\Q)$) is entirely within 
the level-3 component $V_{\alpha_3}^{(3)}$ or not. 
} %
Hodge-isomorphism classes are 
labeled by $\alpha \in {\cal A}^k(M)$. First, whenever $V_a^{(k)}$ of one 
$a \in \alpha \in {\cal A}^k(M)$ is of CM-type, all the other $V_{a'}^{(k)}$ with 
$a' \in \alpha$ are of CM-type. So, whether a rational Hodge substructure 
of $H^k(M;\Q)$ is of CM-type or not can be asked for individual Hodge-isomorphism 
classes in ${\cal A}^k(M)$.  Secondly, one can see that the subspace 
$V_\alpha^{(k)} \subset H^k(M;\Q)$ for $\alpha \in {\cal A}^k(M)$ does not 
depend on a choice of a simple substructure decomposition. To see this, 
suppose that there is another simple substructure decomposition, 
$H^k(M;\Q) \cong \oplus_{\beta \in {\cal A}'} \oplus_{b \in \beta} U^{(k)}_b$; 
the set of Hodge-isomorphism classes ${\cal A}'$ should be the same as 
${\cal A}^k(M)$, and the Hodge isomorphism between 
$\oplus_{\alpha} \oplus_{a \in \alpha} V_a^{(k)}$ 
and $\oplus_{\beta} \oplus_{b \in \beta} U_b^{(k)}$ should be block-diagonal 
with respect to $\alpha, \beta \in {\cal A}^k(M)$. The subspace 
$\oplus_{b \in \alpha} U_b^{(k)} \subset H^k(M;\Q)$ is therefore identical to 
$V_\alpha^{(k)}$. As a candidate of a rational Hodge substructure 
$V_M^{(k)} \subset H^k(M;\Q)$ in the condition 4 of Conj.\ 2, it is enough 
to think of the form $\oplus_{\alpha \in {\cal A}^k(M)_{\rm sub}} V_\alpha^{(k)}$, 
with the Hodge-isomorphism classes running over some subset ${\cal A}^k(M)_{\rm sub}$
of ${\cal A}^k(M)$. 

There is a subset ${\cal A}^k_B(M) \subset {\cal A}^k(M)$ characterized 
as the set of those where the image of 
$p^{(k)}_\alpha \circ \pi_M^*: H^k(B;\Q) \rightarrow V_\alpha^{(k)}$ is non-zero; 
$p^{(k)}_\alpha: H^k(M;\Q) \rightarrow V_\alpha^{(k)}$ is the projection. 
The same set of notations (such as ${\cal A}^k(W)$, ${\cal A}^k(W)_{\rm sub}$, ${\cal A}_B^k(W)$) 
is introduced for the SYZ torus fibration $\pi_W: W \rightarrow B$. 
For the purpose of the condition 4(a) in Conj. 2, it is enough 
to choose ${\cal A}^k(M)_{\rm sub} \subset {\cal A}^k(M)$ as large as ${\cal A}^k_B(M)$. 
The conditions 4(a) does not necessarily require that ${\cal A}^k(M)_{\rm sub}$ should 
contain all of ${\cal A}^k_B(M)$, because the images of $H^k(B;\Q)$ in $V_\alpha^{(k)}$'s 
with different $\alpha$'s may be correlated in general. 

Here, we begin with the first test. 
The conditions 4(b) and 4(c) do not treat $M$ and the mirror $W$ democratically, 
at least at first sight. If the Conj. 2 is to provide a necessary and 
sufficient condition for a ${\cal N}=(1,1)$ SCFT to be rational, then 
all of its mirror SCFTs must be rational. So, we have to make sure 
that the condition 4 implies the condition 4[$M\leftrightarrow W$]. 

The condition 4(b)[$M \leftrightarrow W$] follows from the condition 4(a,b,c) in fact. 
To see this, think of the subset ${\cal A}^k(W)_{\rm sub~sub} \subset {\cal A}^k(W)_{\rm sub}$
where the image of $\phi^*$ in the condition 4(c) is non-zero. Then we may replace 
$V_W^{(k)} = \oplus_{\alpha \in {\cal A}^k(W)_{\rm sub}} V_\alpha^{(k)}$ by 
$V_{W{\rm new}}^{(k)} = \oplus_{\alpha \in {\cal A}^k(W)_{\rm sub~sub}} V_\alpha^{(k)}$. 
This rational Hodge substructure of $H^k(W;\Q)$ still satisfies the condition 4(a). 
Because they are in the non-zero image from CM-type simple rational Hodge substructures
(the condition 4(b)), $V_{W{\rm new}}^{(k)}$ is also of CM-type. 

The condition 4(c)[$M\leftrightarrow W$] also follows from the 
conditions 4(a,b,c). Think of the subset ${\cal A}^k(M)_{\rm sub~sub} \subset 
{\cal A}^k(M)_{\rm sub}$ where $\phi^*$ in the condition 4(c) is non-zero. 
Then there is one-to-one correspondence between ${\cal A}^k(M)_{{\rm sub~sub}}$
and ${\cal A}^k(W)_{{\rm sub~sub}}$. We may replace $V_M^{(k)} = 
\oplus_{\alpha \in {\cal A}^k(M)_{\rm sub}} V_\alpha^{(k)}$ by 
$V_{M{\rm new}}^{(k)} = \oplus_{\alpha \in {\cal A}^k(M)_{\rm sub~sub}} V_\alpha^{(k)}$, 
and yet the condition 4(a) is satisfied. Now, we construct a Hodge morphism 
$\psi*: V_{W{\rm new}}^{(k)} \rightarrow V_{M{\rm new}}^{(k)}$ as follows, 
by focusing on each one-to-one correspondence pair 
$\alpha \in {\cal A}^k(M)_{{\rm sub~sub}}$
and $\beta \in {\cal A}^k(W)_{\rm sub~sub}$. 
The vector space $V^{(k)}_\beta \subset H^k(W;\Q)$ over $\Q$ can also be seen as 
a vector space over the CM field $K$ of the simple rational Hodge 
structures in $\alpha$ and $\beta$; choose any decomposition 
$V^{(k)}_\beta = {\rm Im}(\phi^*) \oplus [{\rm Im}(\phi^*)]^c$ 
as a vector space over $K$, and fix it. 
The vector space $V^{(k)}_\alpha$ also has a decomposition 
${\rm Ker}(\phi^*) \oplus [{\rm Ker}(\phi^*)]^c$ as a vector 
space over $K$; choose this decomposition in a way 
$[{\rm Ker}(\phi^*)]^c$ contains the image 
$p^{(k)}_\alpha \circ \pi_M^* (H^k(B;\Q))$.
Then the Hodge morphism $\phi^*: [{\rm Ker}(\phi^*)]^c \longrightarrow {\rm Im}(\phi^*)$
is invertible; the inverse $\psi^*$ may be extended to 
$[{\rm Im}(\phi^*)]^c \subset V^{(k)}_\beta$ by zero, so we have 
a Hodge morphism $\psi^*: V^{(k)}_\beta \rightarrow V^{(k)}_\alpha$. 
This completes the construction of a Hodge morphism $\psi^*: V_{W{\rm new}}^{(k)} \rightarrow
V_{M{\rm new}}^{(k)}$. 
By construction, $\phi^*$ is an isomorphism between the injective images of 
$H^k(B;\Q)$ within $V_{M{\rm new}}^{(k)}$ and $V_{W{\rm new}}^{(k)}$, 
and $\psi^*$ gives the inverse of $\phi^*$ on those injective images. 
So, the condition 4(c)[$M\leftrightarrow W$] follows indeed. 

The other test is to see if Conjecture 2 (gen.) is consistent with 
Conjecture 1 (self-mirror); the latter is more reliable because it has been 
tested by the case $M=T^4$. When we read the conditions 4  of Conjecture 2 for 
a self-mirror manifold $M$, they appear to be weaker than the conditions 4 and 5 of Conj.\ 1.  
This raises a concern that the conditions in Conj.\ 2 might not be 
strong enough to be a sufficient condition for an SCFT to be rational. 

It is beyond the scope of this article to run this test on all 
the self-mirror families, but we can do it on the family $M=T^4$ here. 
To see that the conditions 4 and 5 of Conj.\ 1 
follow from the condition 4 of Conj.\ 2 for $M=T^4$, 
note first that the rational Hodge structure 
$H^1(M;\Q)$ (not necessarily of CM-type) is either (I) simple, or (II) 
can be split into two simple Hodge substructures. In the case (II), 
each of the two simple components of $H^1(M;\Q)$ contains 1-dimensional 
subspace of $\pi_M^*( H^1(B;\Q) )$; in any SYZ mirror, each complex 
coordinate contains both one SYZ fiber direction and one base direction.
This means that $V_M^{(k=1)}$ in the condition 4 of Conj.\ 2 should always be 
the entire $H^1(M;\Q)$, because of the condition 4(a) in Conj.\ 2. 
Repeating the same argument, one finds that 
$V_W^{(k=1)} = H^1(W;\Q)$ ---(*). 
So, the condition 4(b) of Conj.\ 2 implies that the rational Hodge structure 
on the entire $H^1(M;\Q)$ is of CM-type, regardless of the case (I) or (II). 
The Hodge structures on $H^k(T^4;\Q)$ of all other $k$'s are also of CM-type then. 
Secondly, the Hodge morphism $\phi^*: H^1(M;\Q) \rightarrow H^1(W;\Q)$
in the condition 4(c) of Conj.\ 2 cannot have a kernel; even in the case (II), 
the Hodge morphism $\phi^*$ should map the two simple substructures of $H^1(M;\Q)$
to the two simple substructures of $H^1(W;\Q)$ without a kernel, without a cokernel, 
to meet the requirement on $\phi^*$ in the condition 4(c) (remember (*)). 
So, the Hodge morphism $\phi^*$ must be a Hodge isomorphism, as in the 
condition 5 of Conj.\ 1. 

\vspace{5mm}

Here are two remarks. First, note that the first test motivated an idea of 
restricting the range (i.e., $V_M^{(k)} \subset H^k(M;\Q)$) 
in which the Hodge morphism $\phi^*$ is defined. 
We could replace the condition 4(c) by that---4(c')---of a Hodge morphism 
$\phi^*: H^k(M;\Q) \rightarrow H^k(W;\Q)$; we have not found how to prove 
existence of $\psi^*: H^k(W;\Q) \rightarrow H^k(M;\Q)$ by relying 
on the conditions 4(a,b,c), however, in the absence of an extra observation 
in either variations of Hodge structures, SYZ $T^n$-fibration or SCFTs 
(cf.\ footnote \ref{fn:Fermat}). 

Second, there is an alternative to Conjecture 2, which is to replace the 
condition 4(b) by 
\begin{itemize}
\item [4(b')] the rational Hodge structures on the entire 
$H^k(M;\Q)$ and $H^k(W;\Q)$ are of CM-type. 
\end{itemize}
The alternative version, Conjecture 2' also passes the first test 
(as well as the second one). 
The conditions 4(b) and 4(b') are likely to be different. 
The Borcea--Voisin orbifold Calabi--Yau threefolds will be good 
test cases in finding out which is right  
(cf.\ footnote 6 of \cite{Kidambi:2022wvh}).

\subsection*{Acknowledgments}

The authors thank Abhiram Kidambi for discussions. 
This work was supported in part by 
FoPM, WINGS Program of the University of Tokyo (MO), 
a Grant-in-Aid for Scientific Research on 
Innovative Areas 6003 and the WPI program, MEXT, Japan.  

\appendix

\section{Isogenous Tori Have Isogenous Geometric SYZ-mirrors}
\label{app:isog}



We will prove that isogenous complex tori also have isogenous geometric SYZ-mirrors.
The precise statement is as follows.
Suppose that two complex tori $(T^{2n}_1;I_1)$ and $(T^{2n}_2;I_2)$ are isogenous, and let $\psi:(T^{2n}_1;I_1)\to(T^{2n}_2;I_2)$ be an isogeny. 
Suppose further that the ${\cal N}=(2,2)$ SCFT of a set of data 
$(T^{2n}_2; G_2, B_2; I_2)$ has a geometric SYZ-mirror with the 
$(T^{2n}_{2\circ}1;G_{2}^\circ,B_{2}^\circ;I_2^\circ)$ for the T-dual\footnote{
To be precise, there is no such notion as ``the'' T-dual of a torus-target (S)CFT {\it along} a set of 1-cycles 
$\Gamma_f \subset H_1(T^{2n};\Z)$; we also have to specify the directions $\Gamma_b \subset H_1(T^{2n};\Z)$ 
in which the T-dual is not taken, to specify the dual (S)CFT. We claim that the statement here holds true 
for any $(\Gamma_f^{(2)}, \Gamma_b^{(2)})$ and $(\Gamma_f^{(1)}, \Gamma_b^{(1)})$, for the corresponding 
T-dualities, and for the corresponding sets of geometric data $(T^{2n}_{2\circ}, I_2^\circ)$ 
and $(T^{2n}_{1\circ}, I_1^\circ)$. 
} 
along\footnote{
In section \ref{subsec:Aprime-1-7}, we use this statement for $T^4_1 = T^4$ and 
$T^4_2 = E_1 \times E_2$, and $\Gamma_f^{(2)}\otimes\Q = {\rm Span}_\Q\{ \psi_*(c), \psi_*(d) \}$. 
``The'' T-dual of $E_1 \times E_2$ is $E_2^\circ \times E_1^\circ$ when we choose $\Gamma_b^{(2)}$
so that it is rank-1 in $H_1(E_2;\Z)$ and also rank-1 in $H_1(E_1;\Z)$; this choice is always 
possible because $\Gamma_f^{(1)} \otimes \Q = \Gamma_{f\Q}$ is of the form (\ref{eq:Aprime-GammafQ}).
} %
a rank-$n$ primitive subgroup $\Gamma_f^{(2)}\subset H_1(T^{2n}_2;\Z)$.  
Then we claim that the set of geometric data 
\begin{align}
(T^{2n}_1;G_1,B_1;I_1) \quad \text{where} \quad G_1:=\psi^\ast(G_2),\ B_1:=\psi^\ast(B_2)
\label{eq:def-of-G1-B1}
\end{align}
also has a geometric SYZ-mirror when the T-duality is taken along 
along $\Gamma_f^{(1)}{:=\,}\psi_\ast^{-1}(\Gamma_f^{(2)}\otimes\Q)\cap H_1(T^{2n}_1;\Z)$; the geometric data for the mirror is denoted by 
$(T^{2n}_{1\circ};G_1^\circ,B_1^\circ;I_1^\circ)$. Furthermore, the mirror 
complex tori $(T^{2n}_{1\circ};I_1^\circ)$ and $(T^{2n}_{2\circ};I_2^\circ)$ 
are isogenous.
The rest of this appendix is very elementary and straightforward; we leave this note in this preprint
because this might still save a little bit of time of some readers.

Let us first show that $(T^{2n}_1;G_1,B_1;I_1)$ has a geometric SYZ-mirror for the T-dual along $\Gamma_f^{(1)}$.
That can be done by checking the 
properties that \cite[Prop.\ 8]{VanEnckevort:2003qc} 
\begin{align}
\omega_1|_{\Gamma_f^{(1)}\otimes\R}=B_1|_{\Gamma_f^{(1)}\otimes\R}=0,
\label{eq:SYZ-cond-for-1}
\end{align}
where $\omega_1({-},{-}):=\frac{1}{2}G_1(I_1{-},{-})$ is the K\"{a}hler form.
In fact, (\ref{eq:SYZ-cond-for-1}) immediately follows from the existence of a geometric SYZ-mirror for $(T^{2n}_2;G_2,B_2;I_2)$ (i.e., 
$\omega_2|_{\Gamma_f^{(2)}\otimes\R}=0$ and $B_2|_{\Gamma_f^{(2)}\otimes\R}=0$), and
the fact that 
\begin{align}
&B_1(-,-)=\psi^\ast(B_2)(-,-)=B_2(\psi_\ast-,\psi_\ast-),
\label{eq:B-pullback}\\
&\omega_1(-,-)=\frac{1}{2}G_2(\psi_\ast I_1-,\psi_\ast-)=\frac{1}{2}G_2(I_2\psi_\ast-,\psi_\ast-)=\omega_2(\psi_\ast-,\psi_\ast-);
\label{eq:omega-pullback}
\end{align}
in the latter, we used the fact that the isogeny $\psi$ is a holomorphic 
map and hence
\begin{align}
\psi_\ast I_1=I_2\psi_\ast.
\label{eq:isog-preserve-cpx-struc}
\end{align}

Let us move on to the proof of the existence of an isogeny $\psi^\circ:(T^{2n}_{1\circ};I_1^\circ)\to(T^{2n}_{2\circ};I_2^\circ)$.
That is (see footnote \ref{fn:isog-Hdg-isom}) to construct a linear isomorphism $\psi^\circ_\ast:H_1(T^{2n}_{1\circ};\Q)\to H_1(T^{2n}_{2\circ};\Q)$ satisfying
\begin{align}
\psi_\ast^\circ I_1^\circ=I_2^\circ\psi_\ast^\circ,
\label{eq:mirror-isog-preserve-cpx-struc}
\end{align}
where $\psi^\circ_\ast\otimes\R:H_1(T^{2n}_{1\circ};\R)\to H_1(T^{2n}_{2\circ};\R)$ is also denoted simply by $\psi^\circ_\ast$.

To do so, let us recall how the complex structure $I_i^\circ$ of the mirror is determined by $\omega_i$ and $B_i$ 
$(i=1,2)$ (see \cite[\S\S 2.4--2.5]{VanEnckevort:2003qc} and references therein).
First, the horizontal and vertical generalized complex structures $\cI_i$ and $\cJ_i$ are linear operators 
on $H_1(T^{2n}_i;\R)\oplus H^1(T^{2n}_i;\R)$, 
\begin{align}
&\cI_i:
\begin{pmatrix}
\partial_{X^I} & dX^I
\end{pmatrix}
\longmapsto
\begin{pmatrix}
\partial_{X^J} & dX^J
\end{pmatrix}
\begin{pmatrix}
(I_i)^J_{\ I} & 0\\
(B_i)_{JK}(I_i)^K_{\ I}+(I_i^\transpose)_J^{\ K}(B_i)_{KI} & -(I_i^\transpose)_J^{\ I}
\end{pmatrix},
\label{eq:def-of-generalized-I}\\
&\cJ_i:
\begin{pmatrix}
\partial_{X^I} & dX^I
\end{pmatrix}
\longmapsto
\begin{pmatrix}
\partial_{X^J} & dX^J
\end{pmatrix}
\begin{pmatrix}
(\omega_i^{-1})^{JK}(B_i)_{KI} & -(\omega_i^{-1})^{JI}\\
(\omega_i)_{JI}+(B_i)_{JK}(\omega_i^{-1})^{KL}(B_i)_{LI} & -(B_i)_{JK}(\omega_i^{-1})^{KI}
\end{pmatrix}.
\label{eq:def-of-generalized-J}
\end{align}
Second, the winding and Kaluza--Klein charges of the SYZ-mirror pairs of SCFTs  
are identified through the lattice isometry  
\begin{align}
&g:H_1(T^{2n}_i;\Z)\oplus H^1(T^{2n}_i;\Z)\to H_1(T^{2n}_{i\circ};\Z)\oplus H^1(T^{2n}_{i\circ};\Z) \quad \text{such that}\\
&g(\Gamma_b^{(i)}\oplus(\Gamma_f^{(i)})^\vee)=H_1(T^{2n}_{i\circ};\Z),\quad g((\Gamma_b^{(i)})^\vee\oplus\Gamma_f^{(i)})=H^1(T^{2n}_{i\circ};\Z);
\end{align}
its matrix form is 
\begin{align}
g=
\begin{pmatrix}
\bm{1} & 0 & 0 & 0\\
0 & 0 & 0 & \bm{1}\\
0 & 0 & \bm{1} & 0\\
0 & \bm{1} & 0 & 0
\end{pmatrix}
\end{align}
when presented in the basis of $\Gamma_b^{(i)}\oplus\Gamma_f^{(i)}\oplus(\Gamma_b^{(i)})^\vee\oplus(\Gamma_f^{(i)})^\vee$
(cf.\ footnote \ref{fn:construction-of-mirror}). 
Now, the horizontal generalized complex structure of the mirror 
(i.e., ${\cal I}^\circ$) is given by the 
vertical generalized complex structure before the mirror (i.e., ${\cal J}$) 
through 
\begin{align}
\cI_i^\circ=g\cJ_ig^{-1}.
\label{eq:mirror-cpx-struc}
\end{align}
The complex structure $I_i^\circ$ of the mirror is extracted from the upper-left $2n\times2n$ block 
of ${\cal I}_i^\circ$, which is further given in terms of $\omega_i$ and $B_i$.

A relation between the complex structures $I_i^{\circ}$ ($i=1,2$) of the mirrors can be derived from 
the relations between the $\omega_i$'s and $B_i$'s of the isogenous pair before the mirror. 
The relations between the latter,  (\ref{eq:B-pullback}) and (\ref{eq:omega-pullback}), are written in 
the matrix form as
\begin{align}
(B_1)_{JK}=(\psi_\ast^\transpose)_J^{\ J'}(B_2)_{J'K'}(\psi_\ast)^{K'}_{\ K} \quad \text{and} \quad (\omega_1)_{JK}=(\psi_\ast^\transpose)_J^{\ J'}(\omega_2)_{J'K'}(\psi_\ast)^{K'}_{\ K}.
\end{align}
Therefore, we have
\begin{align}
\Psi\cJ_1=\cJ_2\Psi, \quad \text{where} \quad \Psi:=\begin{pmatrix}
(\psi_\ast)^{J'}_{\ J} & 0\\
0 & (\psi_\ast^{\transpose-1})_{J'}^{\ J}
\end{pmatrix}.
\end{align}
This means, because of (\ref{eq:mirror-cpx-struc}), that 
\begin{align}
g\Psi g^{-1}\cI_1^\circ=\cI_2^\circ g\Psi g^{-1}.
\label{eq:Psi-preserve-generalized-I}
\end{align}

The $4n \times 4n$ matrix $g\Psi g^{-1}$ is not a general $\Q$-valued matrix, but has a structure. 
Since $\Gamma_f^{(1)}$ was defined to be $\psi_\ast^{-1}(\Gamma_f^{(2)}\otimes\Q)\cap H_1(T^{2n}_1;\Z)$, the isogeny $\psi_\ast:H_1(T^{2n}_1;\Q)\to H_1(T^{2n}_2;\Q)$ has the following structure:
\begin{align}
\psi_\ast
\begin{pmatrix}
\Gamma_b^{(1)} & \Gamma_f^{(1)}
\end{pmatrix}
=
\begin{pmatrix}
\Gamma_b^{(2)} & \Gamma_f^{(2)}
\end{pmatrix}
\begin{pmatrix}
P_{11} & 0\\
P_{21} & P_{22}
\end{pmatrix}.
\end{align}
As a result, the upper-right $2n \times 2n$ block of 
\begin{align}
g\Psi g^{-1}=\begin{pmatrix}
P_{11} & 0 & 0 & 0\\
0 & P_{22}^{\transpose-1} & 0 & 0\\
0 & -(P_{22}^{-1}P_{21}P_{11}^{-1})^\transpose & P_{11}^{\transpose-1} & 0\\
P_{21} & 0 & 0 & P_{22}
\end{pmatrix}
\end{align}
vanishes. 

We are now ready to read out from the upper-left $2n\times2n$ block of the equation (\ref{eq:Psi-preserve-generalized-I}) 
that 
\begin{align}
\psi^\circ_\ast:=
\begin{pmatrix}
P_{11} & 0\\
0 & P_{22}^{\transpose-1}
\end{pmatrix}: \; H_1(T^{2n}_{1\circ};\Q)\to H_1(T^{2n}_{2\circ};\Q)
\end{align}
is a Hodge isomorphism (i.e., the condition (\ref{eq:mirror-isog-preserve-cpx-struc}) is satisfied).
Isogenies $T^{2n}_{1\circ} \longrightarrow T^{2n}_{2\circ}$ are obtained as appropriate integer 
multiples of $\psi_*^\circ$ above (cf.\ footnote \ref{fn:isog-Hdg-isom}).

\bibliographystyle{unsrt}
\bibliography{complexMultiplication}

\end{document}